# Spin-phonon interactions revisited: far-infrared emission, Raman scattering, and high-resolution X-ray diffraction at the Néel temperature in LaFeO$_3$

Néstor E. Massa,[1,*/]  Javier Gainza [2]  Aurélien Canizarès,[3] Leire del Campo,[3]

and

José Antonio Alonso.[4]

[1] Centro CEQUINOR, Consejo Nacional de Investigaciones Científicas y Técnicas, Universidad Nacional de La Plata, Blvd. 120 1465, B1904 La Plata, Argentina\

[2] European Synchrotron Radiation Facility (ESRF), 71 Avenue des Martyrs, 38000 Grenoble, France.

[3] Centre National de la Recherche Scientifique, CEMHTI UPR3079, Université Orléans, F-45071 Orléans, France

[4] Instituto de Ciencia de Materiales de Madrid, CSIC, Cantoblanco, E-28049 Madrid, Spain

:Corresponding author:
*Néstor E. Massa,  e-mail: neemmassa@gmail.com

**Physics Subject Headings (PhySH)**

**Disciplines**

- Condensed Matter, Materials & Applied Physics

Research Areas
. Antiferromagnetism . Canted ferromagnetism Magnetic anisotropy. Chemical bonding. Optical phonons. Structural properties.

Physical Systems
$LaFeO_3$ Ferrites. Multiferroics. Strongly correlated systems. Perovskites.

Techniques
High-resolution X-ray diffraction, Raman Scattering, Fourier transform infrared emission spectroscopy.,

# Abstract

We report on the structural evolution, spin-phonon interactions, and magnetoelastic effects in bulk LaFeO$_3$ perovskite across its antiferromagnetic transition. Our investigation is motivated by anomalies in the zone-center phonon bandwidth and peak positions, which, as superexchange precursors, deviate in far-infrared emission and high-temperature Raman scattering spectra from the behavior expected in a purely anharmonic lattice. High-resolution synchrotron X-ray diffraction (SXRD) patterns provide new and intriguing insights into the interplay between the perovskite lattice and the magnetic transition.

We found that while the lattice constants exhibit slight deviations at the Néel temperature ($T_N$), due to spin-phonon interactions, individual interatomic bonds display distinct behavior. The octahedral B-site basal plane, involving oxygen ion pairs that mediate superexchange interactions, displays bond contraction (or elongation) assimilable to the observed in X-ray diffraction for negative thermal expansion. These bond deviations reach a singular point near $T_N$, where a sharp change in bond length is observed. A similar abrupt change is also detected in the nearest-neighbor La–O2 basal plane distances, which we interpret as arising from ion size differences within the 3d–2p hybridized states associated with the superexchange mechanism

High-resolution SXRD measurements of the c-axis $Fe^{3+}$–O1 and $La^{3+}$–O1 apex distances reveal a distinct inflection at the Néel temperature. This behavior supports the role of lattice distortions and crystal field modifications in the emergence of non-collinear weak ferromagnetism in RFeO$_3$ (R = rare earth) perovskites.
Furthermore, our SXRD patterns show an increase in diffraction line intensities peaking at $T_N$, suggesting a straightforward indicator for spin-phonon interactions that also resolves into lattice metastability deduced by nonlinear thermal expansion in the paramagnetic phase. Preliminary results across the full RFeO$_3$ (R = rare earth) series indicate a shared structural framework among all members, potentially offering insights into previously inconclusive structural interpretations in oxides with common octahedral sublattices.

# Introduction

The perovskite-distorted ABO₃ family of compounds serves as a paradigm in the search for materials where structural, magnetic, and electronic modifications can drive fundamental transformations. Manipulating these properties may lead to the realization of ideal magnetoelectric multiferroics with enhanced tunability of magnetic or electric ordering. The lattice is composed of two types of polyhedra: BO$_6$ octahedra and AO$_{12}$ dodecahedra, where B is a transition metal and A is a rare earth element. These polyhedra form interconnected sublattices in a crystal structure described by the cubic Pm-3m space group. This high-temperature arrangement undergoes structural distortions upon cooling, primarily due to octahedral tilting and dodecahedral deformation, both of which depend on the size of the rare earth ion. As a result, the symmetry deviates from cubic, and most perovskites adopt rhombohedral or orthorhombic space groups with four formula units per unit cell and centrosymmetry, inherently excluding polar ferroelectric ion displacements.

Among these transition-metal distorted perovskites, the RFeO$_3$ (R = rare earth) series is one of the most extensively studied.[1] Within this series, LaFeO$_3$ has been proposed to crystallize in a simple-cubic perovskite structure (space group Pm-3m) above ~2140 K [2]. Upon cooling, it undergoes a centrosymmetric rhombohedral distortion caused by the rotation of adjacent FeO$_6$ octahedra in opposite directions. At approximately T$_{Rh}$ ~1228 K, LaFeO$_3$ transforms into an orthorhombic phase (space group Pbnm, D$_{2h}^{16}$) [3, 4, 5] following an out-of-phase rotation of FeO$_6$ octahedra about the a- and b-axes and in-phase rotations about the c-axis. As a Mott insulator with the lightest and largest rare earth ion, LaFeO$_3$ is known to be the least orthorhombically distorted compound in the RFeO$_3$ series. The atomic positions in the orthorhombic phase at ambient temperature have been reported earlier by Marezio et al. [6,7,8]

Magnetic ordering in the RFeO$_3$ series occurs below the Néel temperature (T$_N$), which ranges from ~740 K for LaFeO$_3$ to ~623 K for LuFeO$_3$ [2, 9,10] In LaFeO$_3$ below ~740 K, the high-spin Fe$^{3+}$ ion (t$_{2g}^3$e$_g^2$, S = 5/2) gives rise to Heisenberg-type exchange interactions[11,12] in the form of Fe$^{3+}$–O–Fe$^{3+}$ superexchange pathways (with a bond angle of ~157°), resulting in a G-type antiferromagnetic structure in the Γ$_4$ (G$_x$, A$_y$, F$_z$) configuration[13] This corresponds to antiferromagnetic ordering along the a-axis, weak A-type antiferromagnetism along the b-axis, and canted ferromagnetism along the c-axis. The non-colinear arrangement leads to canting of the Fe$^{3+}$ spins resulting in a weak net magnetization yielding fundamental magnon spectra (spin waves) detectable in terahertz spectroscopy even though the spin structure remains primarily antiferromagnetic. [14] opening the possibility of using an electric-field control of magnetization for the design of magnetoelectric or spintronic devices. $LaFeO_3$ has also been reported to exhibit ferroelastic behavior[15]

Here we present a detailed analysis of the temperature-dependent role of the lattice, focusing on local bond length behavior at the transition to G-type antiferromagnetic ordering in bulk $LaFeO_3$. This transition occurs at temperatures marked by classical reorientations attributed to B-site octahedral tilting and the consequent deformation of A-site polyhedra. Upon cooling, the structure gradually evolves below ~1228 K into a mildly distorted orthorhombic paramagnetic lattice which transitions at the Néel temperature ($T_N$) as pointed into a $G_x$ antiferromagnetic phase along the a-axis and a non-collinear $F_z$ ferromagnetic one along the c-axis.

We analyze high-resolution synchrotron radiation X-ray diffraction (SXRD) patterns (Fig. S1)[16], from 1173 K down to room temperature complemented by observations of zone-center phonon behavior from our high-temperature far-infrared emission (Fig. S2)[16] and Raman scattering (Fig. S3)[16] measurements. Between 1173 K and 740 K, only subtle changes in bond lengths are observed, suggesting a gradual decrease in octahedral relaxation. However, this trend is disrupted by a sharp, resonant-like change in the octahedral basal distances at $T_N$, suggesting that the macroscopic volume modification relieves the built-in strain caused by temperature-dependent spin-lattice interactions in the paramagnetic state. The inferred orbital readjustments appear as dynamically fluctuating bond displacements, deduced from the orthorhombic metric in our high-resolution diffraction pattern analysis.

We find that both lattice and internal sublattice distortions significantly impact exchange interactions below the Néel temperature. Spin-phonon coupling relaxes gradually in a thermally driven dynamic regime, showing alternating bond contractions and elongations in the octahedra. This culminates in a distinctive splitting of the Raman stretching phonon band near 650 K, reminiscent of orbital ordering in a truly orthorhombic Pbnm ($D_{2h}^{16}$) environment. In this temperature range, local positive and negative bond expansions are observed as correlated basal-plane pairs among the nine La–O2 nearest-neighbor distances in the twelve-fold coordinated $O^{2-}$ polyhedral configuration. The temperature-dependent increase in diffraction line intensity, peaking at $T_N\_$, distinctly corroborates a straightforward method for detecting spin-phonon interactions

Our results also help resolve previously reported discrepancies regarding the non-collinear spin canting associated to *Fz* weak ferromagnetism . LaFeO$_3$ exhibits out-of-plane deviations of less than 1° . These deviations, which have been the source of a long debate, are associated with the strong nearest neighbor oxygen bridging exchange interaction and the Fe$^{3+}$ large magnetic moment in tilted octahedra.. Our observation of an unambiguous slope break at $T_N$ in the thermal evolution of Fe-O1 and La-O1 distances supports the RFeO$_3$ (R= rare earth) compounds induced lattice anisotropies determining the non-collinear driving mechanism[17] and against the attributed Dzyaloshinskii-Moriya antisymmetric spin exchanges[18. 19]

# .Experimental details

Standard ceramic synthesis procedures have been used to prepare polycrystalline samples of LaFeO$_3$. Stoichiometric amounts of analytical-grade Fe$_2$O$_3$ and La$_2$O$_3$ oxide powders were thoroughly ground and heated in air at 1000 °C for 12 h, followed by heating at 1300 °C for 12 h in alumina crucibles. Pellets ~ 1 cm in diameter and less than 2 mm thick were prepared by uniaxial pressing of the raw powders and sintering the resulting disks at 1300 °C for 2 h. The purity of the samples was verified by X-ray powder diffraction (XRD) conducted at room temperature using Cu-Kα radiation.

Far- and mid-infrared near-normal emission from LaFeO$_3$ was measured with 2 cm$^{-1}$ resolution over the temperature range of 370 K to 1800 K using two Fourier transform infrared (FTIR) spectrometers. A dual-interferometer setup—comprising a Bruker Vertex 80v and a Bruker Vertex 70—coupled to a rotating table housed inside a dry-air enclosure enabled simultaneous measurement of spectral emittance across two distinct spectral ranges from 40 cm$^{-1}$ to 6000 cm$^{-1}$. The sample, heated using a 500 W pulsed Coherent CO$_2$ laser, was positioned on the rotating table at the focal point of both spectrometers, in a location equivalent to that of the internal radiation sources within the instruments. In this measurement configuration, the sample, placed outside the spectrometers, acts as the infrared radiation source, while the internal sample chambers remain empty. The mid-infrared region (500 to 6000 cm$^{-1}$) was detected using a DTGS detector, and the far-infrared vibrational region (40 to 800 cm$^{-1}$) was covered using a helium-cooled bolometer.

Once the optical data were acquired emittance spectra were calculated using the three measured interferograms,

$$E(\omega, T) = \frac{FT(I_S - I_{RT})}{FT(I_{BB} - I_{RT})} \times \frac{P(T_{BB}) - P(T_{RT})}{P(T_B) - P(T_{RT})} E_{BB} \qquad (1)$$

where $FT$ stands for Fourier Transform, and $I$ for measured interferogram i.e., sample, $I_s$; black body, $I_{BB}$; and, environment, $I_{RT}$. Here, $P$ is the Planck's function taken at different temperatures T; i.e., sample, $T_S$; blackbody, $T_{BB}$; and surroundings, $T_{RT}$. $E_{BB}$ is a correction that corresponds to the normal spectral emissivity of the black body furnace reference (a $LaCrO_3$ Pyrox PY 8 commercial oven) and considers its non-ideality[20] We then place our spectra in a more familiar near-normal reflectivity framework using the second Kirchhoff law, that is,

$$R = 1 - E \qquad (2)$$

where $R$ is the sample reflectivity. This relation assumes that in the spectral range of interest any possible transmission is negligible, emissivity just being the complement of reflectivity.

This allows computing oscillator frequencies using a standard multioscillator dielectric simulation with the dielectric function $\varepsilon(\omega)$, given by

$$\varepsilon(\omega) = \varepsilon_1(\omega) - i\varepsilon_2(\omega) = \varepsilon_\infty \prod_j \frac{(\omega_{jLO}^2 - \omega^2 + i\gamma_{jLO}\omega)}{(\omega_{jTO}^2 - \omega^2 + i\gamma_{jTO}\omega)} \qquad (3)$$

$\varepsilon_\infty$ is the high-frequency dielectric constant taking into account electronic contributions; $\omega_{jTO}$ and $\omega_{jLO}$, are the transverse and longitudinal optical mode frequencies, and $\gamma_{jTO}$ and $\gamma_{jLO}$ their respective damping.

The real $(\varepsilon_1(\omega))$ and imaginary $(\varepsilon_2(\omega))$ parts of the dielectric function (complex permittivity, $\varepsilon^*(\omega)$) is then estimated from fitting the data using the reflectivity R given by[19]

$$R(\omega) = \left| \frac{\sqrt{\varepsilon^*(\omega)} - 1}{\sqrt{\varepsilon^*(\omega)} + 1} \right|^2 \qquad (4)$$

High-temperature inelastic backscattering Raman spectra from polycrystalline samples were recorded at 1 cm$^{-1}$ resolution with the 632.9 nm He-Ne excitation laser line using a Qontor l spectrometer equipped with a Leica 100xm microscope focusing objective (NA 0.85) with a laser spot size of ~ 1 $\mu m$ in diameter. Laser power on the sample was 1 mW. Sample temperature variation was achieved using a Linkam 600 heating stage. The measurements were limited up to 800 K at which temperature the background thermal emission started overcoming phonon profiles

All optical measurements have been done in increasing temperature runs.

High-resolution synchrotron X-ray diffraction (SXRD, $\lambda \approx 0.35418$ Å) measurements were performed at different temperatures at the ID22 beamline of the European Synchrotron Radiation Facility (ESRF, Grenoble)[21] . The sample was contained in a 0.5 mm diameter quartz capillary that was rotating during the data acquisition at 991 rpm. High-temperature data were collected one after the other on the same spot of the capillary using a hot air blower, with temperatures reaching up to ~1173 K . Diffraction patterns. were recorded in 100 K intervals, with finer steps of 50 K around the Néel temperature ($T_n$). All X-ray patterns were (Fig. S1) subsequently analyzed using Rietveld refinements[22] and the structural parameters were obtained with the FullProf software. These refinements yielded bulk lattice parameters, atomic positions, and displacement factors, assuming the space group *Pbnm* ($D_{2h}^{16}$) at each temperature studied.

## RESULTS AND DISCUSSION

One of the major challenges in studying correlated materials is understanding the role magnetic order plays when embedded within the surrounding lattice environment. The relevance of spin-phonon interactions will determine the real magnetization dynamics in the overall properties after octahedral tilting and libration, delineating lattice distortions encompassing elementary excitations. These anisotropic contributions lead to a variety of temperature-dependent features that can be investigated using structural and optical probes.

In the following sections, we present Raman and infrared data and compare them with synchrotron-based high-resolution X-ray diffraction patterns to investigate the intrinsic bulk properties of LaFeO$_3$. We also argue for more in-depth measurements and analyses to gain a more accurate understanding beyond the current picture of this iconic oxide.

Figure 1 shows the SXRD patterns of LaFeO$_3$ perovskite at room temperature (298 K) and at 1173 K, along with their best Rietveld refinements. The results reveal an orthorhombic crystalline phase belonging to the *Pbnm* (D$_2$h$^{16}$) space group at both temperatures. The insets highlights the quality of the profile fitting at high diffraction angles, further supporting this structural assignment

a) *Far infrared emission and Raman scattering zone center phonons*

Among the intriguing properties of distorted perovskites, one often observes alterations in phonon profiles and peak positions associated with magnetic spin ordering. These changes can extend beyond those induced solely by lattice effects and may reflect the strength of spin-phonon interactions emerging near the spin ordering (Néel) temperature. To investigate this, far-infrared and Raman scattering spectroscopies, both well-established detection techniques, have been employed to study the high-temperature magnetic and structural changes of LaFeO$_3$. around its Néel temperature

Our spectra encompass the low-frequency vibrational features corresponding to La–O lattice modes, the mid-frequency bands associated with Fe–O bending vibrations, and the high-frequency modes linked to octahedral stretching and breathing modes in the orthorhombic *Pbnm* (D$_2$h$^{16}$) space group with four molecular units per unit cell. Accordingly, the irreducible representations of the zone-center infrared (ungerade) and Raman (gerade) active modes for LaFeO$_3$ are [23]

$$\Gamma_{IR} = 7B1u + 9B2u + 9B3u \qquad (5)$$

$$\Gamma_{Raman} = 7Ag + 7B1g + 5B2g + 5B3g \qquad (6)$$

Fig. 2 and Figs. S2-S4 in the Supplemental Material [16] show the overall spectral evolution of the temperature-dependent far-infrared emission around the Néel temperature (T$_n$), up to 870 K. Notably, ~ 600 K, a subtle yet distinct change appears near the stretching–breathing mode at ~550 cm$^{-1}$, suggesting an additional contribution beyond regular thermal expansion and high-temperature anharmonicity, which typically dominate vibrational band profiles upon heating. This indicates that just above T$_n$, the lattice is dynamically perturbed due to emerging interactions that are not detectable in the paramagnetic phase but gradually manifest below this temperature. In Fig. 2(c), we mimic this behavior by adding an extra oscillator to the spectral fit. It is important to note that this does not imply a structural phase transition, as the lattice retains its orthorhombic *Pbnm* ($D_{2h}^{16}$) symmetry throughout the entire temperature range studied. However, it does suggest deviations from regular phonon behavior as the magnetic transition temperature is approached. This is particularly relevant because of the wide nature of thermal emission phonon linewidths [full-width at half-maximum(FWHM)] and band overlap, which prevents a clear discrimination. This limitation will be addressed in the following discussion on inelastic Raman scattering measurements.

Figs. 3 and 5 show Raman spectra of LaFeO$_3$ collected from 350 K to 800 K, covering the lattice vibrational and internal mode regions, respectively. As in the infrared case, the lattice modes appear at lower frequencies and are separated from the internal modes by weaker features. These weaker bands become Raman-active due to Brillouin zone folding associated with the high-temperature rhombohedral phase. The spectra were deconvoluted using a sum of Gaussian–Lorentzian peaks, in full agreement with numerous previous studies on isostructural compounds [24-26] The active phonon modes were subsequently assigned to the corresponding $A_g$ and $B_g$ species.[27]. Additionally, there have been reports at room temperature of anomalous phonon merging in the spectral regions between 100–200 cm$^{-1}$ and 400–450 cm$^{-1}$. Weber et al. [25] interpreted this behavior as resulting from a coupling between two modes of the same symmetry. In our spectra (arrows in Figures 3 and 5), pairs of phonon bands appear with overlapping intensity profiles and distinct temperature dependencies. As temperature increases, the phonon set centered ~140 cm$^1$ shows a softening component beginning ~450 K. Simultaneously, its counterpart unexpectedly hardens, and the resulting structure disappears ~ 600 K, becoming masked as a shoulder-like broadening in the background of the adjacent phonon profile.

Similarly, bands near 400 cm$^{-1}$ merge at ~ 500 K, matching the intensity of the ~650 cm$^{-1}$ stretching modes (Fig. 5). The latter mode, associated with the O$^{2-}$–Fe$^{3+}$ breathing vibration, is the most intense feature in the spectrum and exhibits splitting upon cooling as in an orbital rearrangement within the orthorhombic *Pbnm* ($D_{2h}^{16}$) space group.

Our main focus, however, is the thermally driven frequency softening as a common feature observed across all vibrational modes. Figs. 4 and 6 show the temperature-dependent peak positions of representative lattice and internal modes. These shifts reflect anharmonic effects, which can be described using the three-phonon model based on second-order cubic anharmonicity written as

$$\omega_{anh} = \omega_0 - C\left(1 + \frac{2}{e^{\frac{\hbar\omega_0}{2kT}}-1}\right) \qquad (7)$$

where $\omega_0$ may be thought of as the 0 K mode frequency (squares on the ordinate axes in column (a) in Figs. 4 and 6), being C an arbitrary slope-dependent constant; $\hbar$ is the reduced Plank's constant and k is the Boltzmann's constant.[28]Examples of curve adjustments using Eq. (7) for the vibrational frequencies are shown in Figs. 4(a) and 6 (a). It is evident that as we approach the Néel temperature $T_N$, there is a distinctive peak frequency deviation in the experimental data that extends to the paramagnetic-canted antiferromagnetic $\Gamma_4$ phase transition. We observed a deviation from the expected thermal trend (Figs. 4b. 6b) along with peak broadening (Figs. 4c, 6c) that exceeds what would be anticipated, even when accounting for four-phonon processes [28]. Overall, it should be noted that phonon widths of vibrational modes in a distorted perovskite lattice are characterized by fluctuating equilibrium positions originating in the lattice metastability that prevents the extraction of a purely anharmonic phonon shift and line width

These deviations from purely thermally driven anharmonic behavior indicate spin–phonon interactions. In insulating $LaFeO_3$, this rules out other contributions, suggesting that the observed perturbation is most likely correlated with magnetostriction. They imply lattice distortions involving changes in bond lengths and sublattice tilting angles as the $FeO_6$ octahedra adjust in response to the antiferromagnetic ordering of Fe spins in neighboring $FeO_2$ planes. A similar effect was previously observed in $LaMnO_3$ [29]

It is also worth noting that, as the temperature decreases, spin ordering becomes more pronounced and anharmonicity is reduced. This leads to the emergence of Raman-active bands with less dynamical broadening, revealing a form of partial orbital reordering (Fig. 4(c)), previously attributed to the La–Fe ionic size mismatch[2] which influences the octahedral tilting angle[30]. This behavior is further reflected in the deconvoluted FWHM of the bands shown in Figs. 4(c) and 6(c), which relate to phonon lifetimes.

Additionally, frequency-modulated substructures emerge (Figs. 4 and 6; columns (b) and (c))) , suggesting subtle changes associated with ongoing processes that will be quantitatively revealed through the high-resolution diffraction patterns discussed in the next section

b) ***High resolution X-ray diffraction***

High-quality, high-resolution X-ray diffraction provides unprecedented detail on the temperature-dependent changes in lattice parameters and the associated bonds involved in spin ordering. It enables a quantitative understanding of the balancing mechanisms by which antiferromagnetism is established and maintained at temperatures below $T_n$. Here, we focus on the changes in bulk LaFeO$_3$ nearest-neighbor bonds that structurally mediate magnetic exchange interactions within a thermally excited lattice [31].

Figure 7 shows the unit cell parameters determined from temperature-dependent SXRD data, refined within the *Pbnm* ($D_{2h}^{16}$) space group. The orthorhombic lattice constants *a, b*, and *c* deviate slightly from a purely linear increase with temperature. This departure may be heightened by performing linear fitting of the antiferromagnetic and paramagnetic phases, as the data points converge at a distinct inflection point intersecting at $T_N$. A closer look reveals that although thermal expansion in the magnetically ordered phase remains approximately linear, there is a need to account for a finite nonlinear contribution, expressed as:

$$\Delta L = c_1 \Delta T + c_2 \Delta T^2 \qquad (8)$$

where $\Delta L$ is the change in length relative to the starting length at $L_0$ and $\Delta T$ is the temperature variation. Upon rearranging, the first term corresponds to the linear coefficient of thermal expansion, defined as $\alpha = \Delta L / L_0 \Delta T$, while the second term $\beta = (\Delta L - c_1 \Delta T)/L_0 (\Delta T^2)$ represents a quadratic deviation.

The thermal expansion coefficient of LaFeO$_3$, calculated over the full temperature range under the linear approximation (Figs. 7 (a, b), and Fig. S5 in the Supplemental Material[16]), is consistent with the data reported by Selbach et al. [2]. The most notable deviation beyond the confidence band is primarily observed in the paramagnetic phase (Fig. S6 in the Supplemental Material[16]), particularly along the c-axis above the Néel temperature (Fig. 7c), where electron–phonon interactions become more relevant.

Nonlinear thermal expansion is mainly attributed to lattice anharmonicities and phase competition, which induce lattice metastability. In LaFeO$_3$, this effect becomes especially prominent as the system approaches the order–disorder phase transition at ~1228 K, where lattice metastability dominates the paramagnetic phase due to the loss of long-range antiferromagnetic correlations. This is also evident in our x-ray diffraction patterns as increased line broadening, acting as a precursor to side-band splitting observed at our highest measured temperatures [Figs. 10(a)–10(c)].

On the other hand, magnetic-related deviations become particularly apparent as a step-like feature at $T_n$, corresponding to the relaxation of internal strain defined as, [32]

$$s=(b-a)/(b+a) \qquad (9)$$

This, shown in Fig. 7 (d), going into the paramagnetic phase, brings up bond-concealed spin-phonon interactions addressed below for $LaFeO_3$ perovskite sublattices B and A .

### i) *B- octahedral sublattice*

Constituting the B-site sublattice of the perovskite structure, $Fe^{3+}$ ions are located at the centers of somewhat distorted $FeO_6$ octahedra. These octahedra become larger and less asymmetrical at the Néel temperature ($T_n$), above which the lattice may be considered metastable with respect to the rhombohedral-to-orthorhombic order–disorder phase transition at ~ 1228 K (Fig. S7 in the Supplemental Material) [16].

The $LaFeO_3$ lattice is built from alternating layers of corner-sharing $FeO_6$ octahedra, yielding Rietveld-refined bond lengths that fall into two sets: Fe–O2 and Fe–O2′ in the __*ab*__ basal plane, and a third, Fe–O1, representing the apical bond along the __*c*__ axis. The temperature dependence of these interatomic distances across $T_n$ is shown in Fig. 8(a,c). We found that these bond lengths exhibit a correlated coupling with the magnetic ordering in the temperature range where the Fe magnetic moments adopt an in-plane *b*-aligned ferromagnetic configuration, while simultaneously establishing antiferromagnetic couplings with six nearest-neighbor $Fe^{3+}$ ions via $Fe^{3+}$–O2–$Fe^{3+}$ superexchange interactions. This behavior corresponds to the well-known $G_x$-type antiferromagnetic ordering, with Fe magnetic moments aligned along the crystallographic __*a-*__axis [33].

Starting from room temperature, our temperature-dependent high-resolution X-ray diffraction patterns provide a more quantitative and detailed view of the temperature evolution of the Fe–O2 and Fe–O2′ bond lengths in the octahedral basal plane . The decrease in Fe–O2′ bond length (dashed line in Fig.8a) with increasing temperature is superposed with a more pronounced spin-phonon anomaly around and at the Néel temperature. This indicates an overall temperature-driven contraction, or negative expansion. In contrast, the Fe–O2 bond in the basal plane exhibits elongation (Fig. 8c), showing a distinct singularity offset at $T_N$. In both cases there is an abrupt bond length change near $T_N$. That orchestrated bond behavior, unambiguously detected thanks to the high angular resolution of our SXRD, can be interpreted as a manifestation of the pivotal swinging mechanism frequently invoked to explain lattice negative expansion in intersublattice distances of bridging-oxygen compounds. Such distortions are discussed in Ref. 34 In this scenario, X-ray-diffraction/measurements give the average values (over the diffraction domain) of bond lengths and lattice constants. These average bond lengths, i.e. the distance between the mean positions of two atoms, would now appear as shorter (longer) distances as the bond is affected by positive or negative fluctuations.. The fluctuation becomes larger as the temperature increases, reducing its *apparent* x-axis projected magnitude (Fig. 8 b). On the other hand, the short-range or local bond lengths, yielding the mean distance between ions at local scale, is the parameter measured from Extended X-ray Absorption Fine Structure (EXAFS) oscillations at synchrotron beamlines. [34, 35]

Closer to the Néel temperature, both distances, Fe-O2' and Fe-O2, are found associated to the superexchange integral through dynamical orbital harmonics, 3d-$Fe^{3+}$ - 2p-$O^{-2}$ hybridization, and spin-phonon interactions. This is also inferred in the octahedra volume dependence at $T_N$. (Fig. 8 (f)) and changes in the *ab* octahedral basal plane (Fig 8 (g))  It is concurrent with fluctuating (non-static) orbitals shift due to electron hopping from 2p–O levels to 3d–Fe ions, giving rise to the $G_x$ and $A_\gamma$ antiferromagnetic order. Then, since it is known that the $Fe^{3+}$ remains in 5/2 high spin through the AFM-PM transition we are bound to associate the abrupt length change in basal bond lengths at $T_N$ to the oxygen ion bridging quasi-rigid sublattices, These sublattices are where transition metal orbitals interact, via the oxygen in-between, with both d orbitals coupling to the same p orbital. Spin up and down may hop from O-p to $Fe^{3+}$-d ions [36, 37]

Those bond singularities also exhibit a weaker correlation with the Fe–O1 bond along the *c-*axis, which corresponds to the direction of the net weak noncollinear ferromagnetism (Fig. 8(d)). The thermal evolution of this bond shows a notable change in slope at $T_n$, which—as discussed in the next section—is also reflected as a subtle inflection point in the La–O1 distances (Fig. 9 a,b) . In this context, Bozorth suggested that the origin of this noncollinear ferromagnetism lies in the $Fe^{3+}$ magnetic moment being slightly tilted by the crystal field, deviating from the ideal Pbnm ($D_{2h}^{16}$) orientation, thereby introducing a unique net magnetic component along the *c*-axis [38.39] This proposition is concomitant with our detection of octahedral volume changes at $T_n$, which would necessarily alter the crystal field environment. Such distortions mediate a deviation of the $Fe^{3+}$ moment from its ideal exchange orientation, producing a net magnetic component along the *c*-axis. Zhou et al [17] have recently pointed out that the linear dependence of the spin canting angle with the rare earth ion size should be taken as evidence of the dominant role of crystal fields in ferromagnetic canting represented in single-ion anisotropy for $RFeO_3$ (R=rare earth) compounds. The slope break observed along the *c*-axis in $LaFeO_3$ (Fig. 8(d)), as well as in our preliminary data from other members of the $RFeO_3$ family, supports this conclusion. The slope change at $T_n$ in the temperature-dependent apical oxygen interatomic distance suggests a distinct crystal field environment in the distorted structure, one capable of inducing ferromagnetism, potentially even more strongly than originally proposed by Bozorth in the early days of crystallography.[38]

This finding also calls into question the dominant role traditionally attributed to spin–orbit coupling and reduces the emphasis on antisymmetric exchange interactions, such as the Dzyaloshinskii–Moriya (DM) interaction, expressed as **D · (S$_i$ × S$_j$)**, where ***D*** is the antisymmetric DM constant and ***S**$_i$*, ***S**$_j$* are the spin moments of neighboring transition metal ion[40] The historical prominence of this interaction is increasingly seen as a consequence of limitations in the resolution of instrumentation available over 60 years ago, which likely obscured subtle structural distortions now known to exist in every distorted perovskite.

## ii)  *A- polyhedron sublattice*

Since the A-site cation lacks dedicated oxygen coordination, changes in the octahedral bond lengths are also reflected in the surrounding dodecahedral environment. In orthorhombic $LaFeO_3$, the ideal cubic 12-fold coordination of the A-site is reduced to nine nearest-neighbor oxygens. Except for the two La–O1 apical bonds, which exhibit only a weak slope change across the Néel temperature (Fig.9 (a, b)), the La–O bonds associated with the basal plane oxygen atoms (O2 and O2′) belong to the nearest octahedral sublattices.. As shown in Fig. 9, the pairs in Fig. 9 (c)-(d); (e)-(f); and (g)-(h), exhibit the same step-like discontinuity at $T_n$ as commented for the $Fe^{3+}$–O bonds in the preceding section. A fourth bond pair lies near the Rietveld refinement cutoff distance and is therefore only detected in our measurements above

~740 K, where the lattice symmetry begins to increase. The same picture may be recreated by choosing the bond origin at O1 and O2 ions(Figs S8, S9 in the Supplemental Material) [16]
That is, anomalous changes in both Fe–O2 and Fe–O2 ′ bond distances are found to be reproduced in all La–O2 bonds as step-like variation amounting at $T_n$ to ~ 0.394 Å↔~1.01 Å length differences linked to factors intrinsic to oxygen ions. Rare earths are not expected to play any role in the combination of direct exchange and electron transfer of the superexchange interaction, while Oxygen plays the role of a mediator of indirect interaction, normalizing the exchange in the magnetic interaction by which spin polarization takes place via p-d hybridization [6]

Those length changes are compatible with a possible Oxygen size modification due to spin state change, in which case spin-down would represent bond length contraction and spin-up would correspond to elongation. Ion radius changes compatible with this have been observed for LaFeO$_3$ at high pressures with the sublattice of Fe$^{3+}$ octahedra in high (S=5/ 2, $^6A_{1g}$) has a 0.078 Å radius, coexisting with those shortened by low-spin (S=1/ 2, $^2T_{2g}$) Fe$^{3+}$ with a 0.068 Å [31, 39] The number of electrons in the $e_g$ states determines the ion size proportional to the thermal population of the spin state [36]

On the other hand, within this scenario, it is also worth noting that a full understanding might also require considering the role of nonlinear onsite and directional polarizability of O$^{2-}$ in the hybridization with the transition metal through dynamical covalency.[45]    Polarizability is the environment temperature-dependent property of orbital overlaps, allowing electron redistribution after a Fe-O2 bond motion. The nonlinear, anisotropic polarizability of the O$^{2-}$ ion arises from its instability as a free ion, which tends toward delocalized oxygen $p$-states. In the lattice, these states hybridize with the $d$-orbitals of neighboring transition metal ions (Fe$^{3+}$). The O$^{2-}$ ion becomes lattice-stabilized through Coulomb interactions with transition metal and rare-earth ions, within the framework defined by different Watson sphere radii [46] [47]

In LaFeO$_3$ this feature would enhance the Fe$^{3+}$–O$^{2-}$–Fe$^{3+}$ antiferromagnetic superexchange interaction through an optimized hybrid entanglement and bond dynamics, leaving also room for noncollinearity. As temperature varies, this process energetically locks in an effective spin–spin interaction established between B-site ions, resulting in antiferromagnetic coupling. Near the Néel temperature, the resonant-like behavior brings RFeO$_3$ (R = rare earth) compounds closer to ferroelectricity, since the interplay between Coulomb and electron–phonon interaction, predominating in the paramagnetic phase, is a known driver leading to ferroelectric instabilities [48] that it is also expected in the multiferroic magnetic and ferroelectric ordering coupling proposed for LaFeO$_3$ [49]
.Additionally, we identified a closely correlated effect in the temperature-dependent X-ray diffraction profiles, which peak at T$_n$ (Fig. 10). In principle, the X-ray scattering amplitude can be described by the sum of elastic and magnetic contributions, as expressed in [50]

$$A_s = r_e \{T + i \frac{\lambda_c}{\lambda} \; [\frac{1}{2} \mathbf{A} \cdot \mathbf{L}(K) + \mathbf{B} \cdot \mathbf{S}(K)]\} \qquad (10)$$

where r$_e$ is the electron radius; T, is the classical term for scattering of radiation by electronic charge only dependent on the radius of a particle of mass m and charge q; $\lambda_c = \frac{\hbar}{mc}$ is the Compton wavelength ($\lambda_c$= 0.0243 Å) with λ the measuring X-ray wavelength. In standard X-ray diffraction $\frac{\lambda_c}{\lambda}$ is of the order of 10$^{-2}$. .. On this, the magnetic scattering contribution to the X-ray diffraction appears between brackets in eq(9).    **A** and **B** are the wave vector and polarization states of the incident and scattered photons and L(*K*) and S(*K*) represent the Fourier transform of angular momentum and spin density, respectively.   The scattering amplitude ratio between both terms is given by  hK/mc in the order of 10$^{-3}$ in diffraction experiments, to which it is necessary to also note that magnetic scattering takes place only for unpaired electrons, making the magnetic contribution negligible for any practical purpose.[48] Our temperature-dependent bulk measurements show diffraction lines distinctively peaking non-resonantly at Néel temperature. This feature may naturally be associated with our discussion on the electron crossover bond imprinted on the superexchange because the change from spin paramagnetic-lattice to spin antiferromagnetic-lattice an electron may be thought to be transferred to an empty level  making the crossover a new channel for the scattering process that adds to the amplitude. The increment in peaking intensity at T$_N$ is found in all measured diffraction lines making superexchange a direct mechanism as  $Fe^{3+}$ ions coupled antiferromagnetically to six $Fe^{3+}$ nearest neighbors via $Fe^{3+}$-$O^{-2}$-$Fe^{3+}$. At the same time, indirectly, it may be thought as a perturbative mechanism propagating coherently throughout the lattice. These findings support the perspective that magnetic coupling in the paramagnetic state is not totally responsible for the magnetic transition, but rather it results from a spin-orbital entangled state. [49] On the other hand, the reduction of magnetic correlations in the paramagnetic phase allows the lattice metastability to be detected in the weakening and broader diffraction profiles. These  eventually split at  our highest temperatures, Figs. 10 (a)-10(c), closer to the rhombohedral structural phase transition $T_{Rh}\sim$ 1228 K (Fig. S7 in Supplemental Material )[16]

# CONCLUSIONS

We investigated the high-resolution thermal structural evolution of LaFeO$_3$ using synchrotron radiation, motivated by deviations from intrinsic anharmonic behavior observed in zone-center phonon frequencies. Superexchange precursors are detected up to a hundred degrees above the Néel temperature. The relatively mild perovskite distortion in this compound enables us to extract information on the coupling of elementary excitations and other effects that may be overlooked in more structurally complex systems.

We found that, while the lattice parameters obtained from standard Rietveld refinement show only weak signatures of the Néel transition—interpreted as indicative of weak spin–phonon coupling—individual molecular bonds behave quite differently. The basal-plane oxygen pairs at the B-site octahedra, which mediate the superexchange interaction, exhibit bond contraction and elongation behaviors reminiscent of the pivotal fluctuations invoked in the interpretation of bulk negative thermal expansion. Both Fe–O2 and Fe–O2′ bond lengths approach critical values near the Néel temperature, reaching a minimum in the Fe–O2′ distance and a complementary maximum in the Fe–O2 distance. Similar abrupt changes in bond lengths are observed at the A-site in the La–O2 bonds lengths.

We hypothesize that these anomalies may be explained by changes in the oxygen ion size due to variations in polarizability and/or spin state within the 3d–2p hybridized states that underpin the superexchange interaction. We also found evidence supporting a role for the lattice in the emergence of weak non-collinear ferromagnetism, consistent with earlier suggestions by Bozorth [38] based on crystal field effects and structural symmetry Both the Fe$^{3+}$ and rare-earth–O1 apex temperature-dependent distances show slope inflections across the Néel temperature, which should be considered a primary lattice contribution alongside spin–orbit coupling in explaining magnetic non-collinearity.

Our high-resolution X-ray diffraction patterns reveal a relative intensity increase in diffraction peaks at the Néel temperature, suggesting a straightforward test for spin–phonon interactions.

This supports the view of superexchange as not merely a local event but a phenomenon coherently propagated through the lattice.

Preliminary results across the full $RFeO_3$ (R = rare earth) family indicate a shared structural and magnetic framework, offering a common blueprint that could help resolve currently inconclusive structural interpretations in other oxides featuring similar corner-sharing octahedral sublattices

## Data availability

The datasets of the high-resolution x-ray diffraction patterns from the current study are available at the ESRF repository at Refs. [50, 51]. Infrared and Raman scattering data are available from the authors upon reasonable request.

## Acknowledgments

JG, JAA, and NEM are indebted to the European Synchrotron Radiation Facility for beamtime allocation in beamline ID22 under Proposal No HC-6056. JAA also acknowledges the financial support of the Spanish "Ministerio de Ciencia e Innovación " (MICINN) through Project Nº PID2021-l22477OB-100.

# Figure Captions

**Figure 1** (color online). X-ray diffraction patterns for LaFeO$_3$ at 298 K and 1173 K in the space group Pbnm (D$_{2h}^{16}$), $Z = 4$. Sharp peaks indicate very good polycrystalline quality. In the Rietveld plot, the red points correspond to the experimental profile and the solid line is the calculated profile, with the differences in the bottom blue line.

**Figure 2** (color online). (a) Near Normal phonon 1-Emissivity of LaFeO$_3$ between 393 K and 813 K. The spectra have been vertically offset for better viewing; (b) Stretching-breathing mode range showing the change in band profile across the Néel temperature, experimental: triangles, full line: fits; (c) Fit oscillators with an extra one ad-hoc (dashed profile) stressing the spin-phonon interaction as the lattice distorts in the magnetic ordered environment.

**Figure 3** (color online) Temperature-dependent LaFeO$_3$ Raman spectra of lattice vibrational modes in the 350 K to 800 K range using laser line λexc = 633 nm. Spectra and fitting traces have been vertically offset for better viewing. Dashed lines indicate temperature-dependent background profiles deconvoluted from the phonon fits.

Figure 4 (color online) (a) Temperature-dependent peak softening of selected modes. Full lines are second-order cubic anharmonicity adjustments after eq(7); (b) temperature-dependent peak mode deviation departing from the purely anharmonic thermally driven; (c) band profile full-width at half-maximum (FWHM) of the same modes. Arrows in columns (b) and (c) point to weak discontinuities related to the antiferromagnetic phase transition at T$_N$ (see text). Squares on the ordinate axes in column (a) are $\boldsymbol{\omega_0}$'s fitting constant thought as a zero Kelvin mode frequency

**Figure 5** (color online) Temperature-dependent LaFeO$_3$ Raman spectra of internal vibrational modes in the 350 K to 800 K range using laser line λexc = 633 nm; spectra and fitting traces have been vertically offset for better viewing.

Figure 6 (color online) (a) Temperature-dependent peak softening of selected modes. Full lines are second-order cubic anharmonicity adjustments after eq(7); (b) temperature-dependent peak mode deviation departing from the purely anharmonic thermally driven; (c) band profile full-width at half-maximum (FWHM) of the same modes. Arrows in columns (b) and (c) point to weak discontinuities related to the antiferromagnetic phase transition at $T_N$ (see text). Squares on the ordinate axes in column (a) are the $\boldsymbol{\omega_0}$'s fitting constant thought as of a 0 K mode frequency

**Figure 7** (color online) Unit cell orthorhombic lattice constants **_a_, _b_,** and **_c_** ; (a) lattice constant **_a_** determined from measured temperature-dependent atomic positions in the space group Pbnm ($D_{2h}^{16}$); linear fits in the antiferromagnetic (full line) and paramagnetic phases (dotted line). Inset: $\alpha_{Gx}$ (K$^{-1}$) linear thermal expansion coefficient in the $G_x$ antiferromagnetic phase, $\alpha_{PARA}$ (K$^{-1}$) linear thermal expansion coefficient in the paramagnetic phase, $\alpha_{linear}$(K$^{-1}$) linear thermal expansion coefficient in the full temperature range, (b) lattice constant **_b_** determined from measured temperature-dependent atomic positions in the space group Pbnm ($D_{2h}^{16}$); and linear fits in the antiferromagnetic (full line) and paramagnetic phases (dotted line). Iinset: $\alpha_{Ay}$(K$^{-1}$) linear thermal expansion coefficient in the antiferromagnetic phase, $\alpha_{PARA}$(K$^{-1}$) linear thermal expansion coefficient in the paramagnetic phase, $\alpha_{linear}$ (K$^{-1}$) linear thermal expansion coefficient in the full temperature range, (c) paramagnetic temperature-dependent lattice constant **_c_** determined from measured atomic positions in the space group Pbnm ($D_{2h}^{16}$); linear (dotted line) and quadratic non-linear (dashed line) fits. Inset: $\alpha_{PARA}$(K$^{-1}$) : linear thermal expansion coefficient in the paramagnetic phase, $\alpha_{nonlinear}$(K$^{-1}$) linear thermal expansion term calculated from the non-linear fit to the lattice constant **_c_** , $\beta_{nonlnear}$(K$^{-2}$) non-linear thermal expansion term calculated from the non-linear fit to the lattice constant **_c_** ; (d) ;attice strain relaxation across $T_N$ after eq. (9) , (d) step-like magnetic-induced deviations at $T_n$, corresponding to the internal strain relaxation [32].

**Figure 8** (color online) (a) Fe-O2' basal-plane bond length (dots) showing the temperature-dependent general trend (dashed line) and the net deviation by spin-phonon interaction about $T_N$; (b) Simplified schematics showing the effect of pivotal libration on an apparent Fe–O bond length (after [34, 35]); (c) Temperature-dependent Fe-O2 basal-plane bond length; (d) Temperature-dependent Fe-O1 oxygen apex bond length showing the distinctive structural anomaly at $T_N$; (e) ) "c" view of the LaFeO$_3$ lattice; (f) Temperature-dependent strain and octahedral volume across $T_N$; (g) Enhance visualization of the temperature-dependent **_ab_** basal plane displayed after a 3.90 Å subtraction in rhombic diagonals calculated from the measured distances shown in (a) and (c).

**Figure 9** (color online) Temperature-dependent A-polyhedron La-O nearest neighbor bond lengths; (a)-(b) La-O1 apical bond lengths, dashed line in (b) is a guide for the eye stressing the breakpoint at $T_N$. (c)-(d), (e)-(f), (g)-(h), (i)-(j) show the different La-O2, La-O2' basal plane bond pairs. Note that the La-O2 in (j) is only considered after the relative increase in symmetry in the paramagnetic phase.

**Figure 10** (color online) Synchrotron high-resolution X-ray diffraction patterns of LaFeO$_3$. (a-) (c) Temperature-dependent x-ray reflections for Miller index [2.0.0], [1,1,2], and [0,2.0] respectively. (d) Temperature evolution of the measured x-ray diffractogram patterns

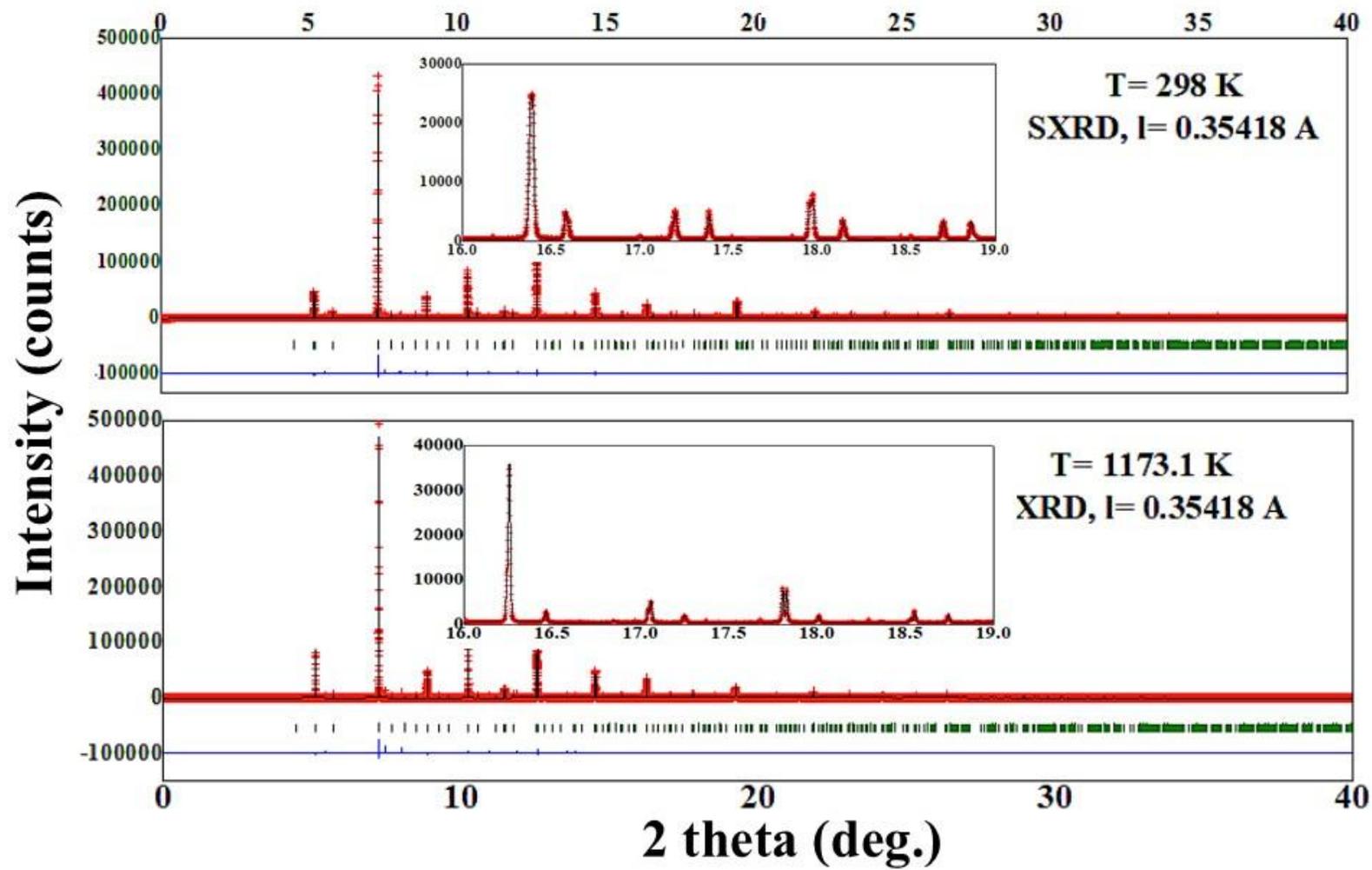

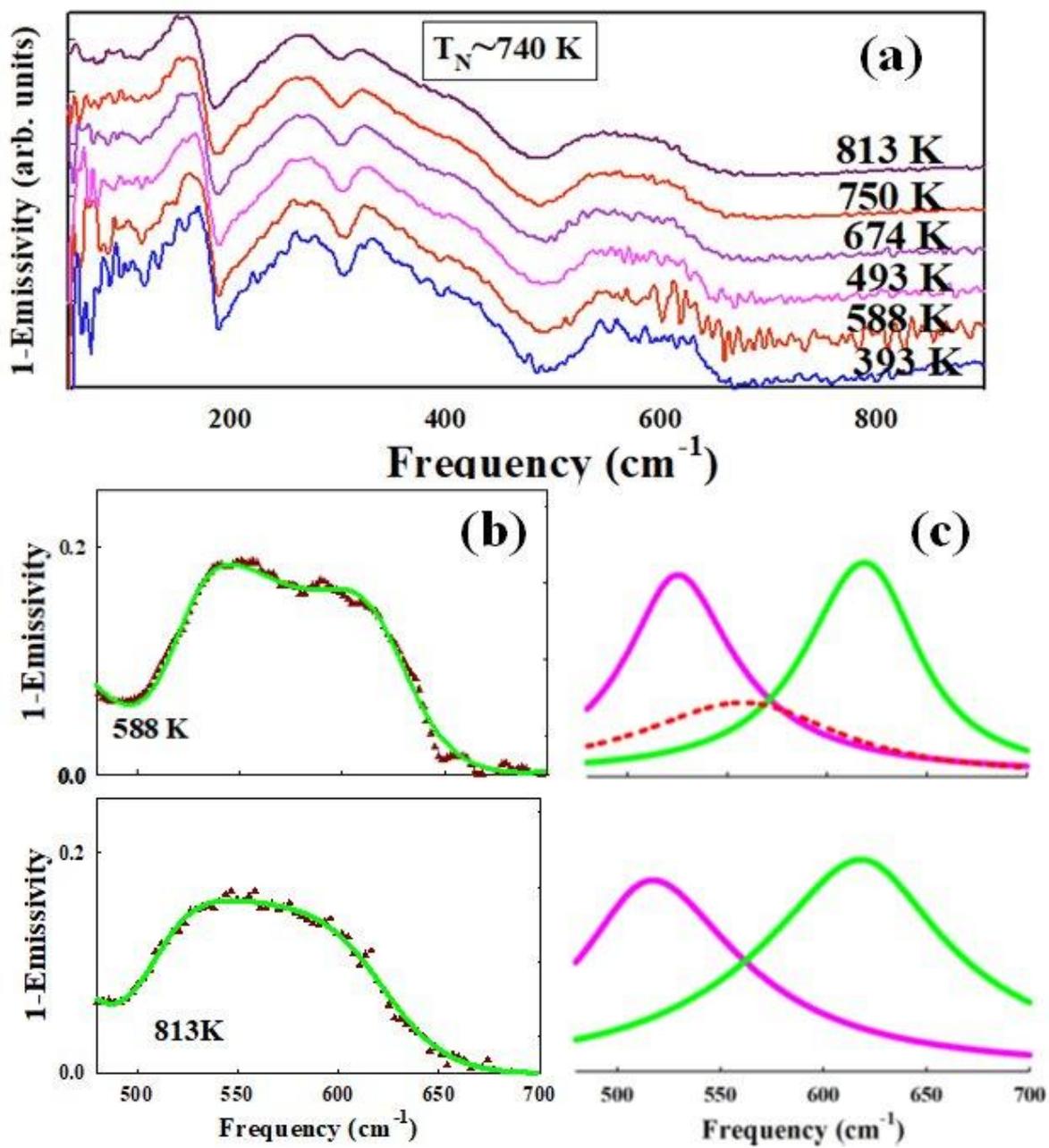

**Fig. 2
MASSA et al**

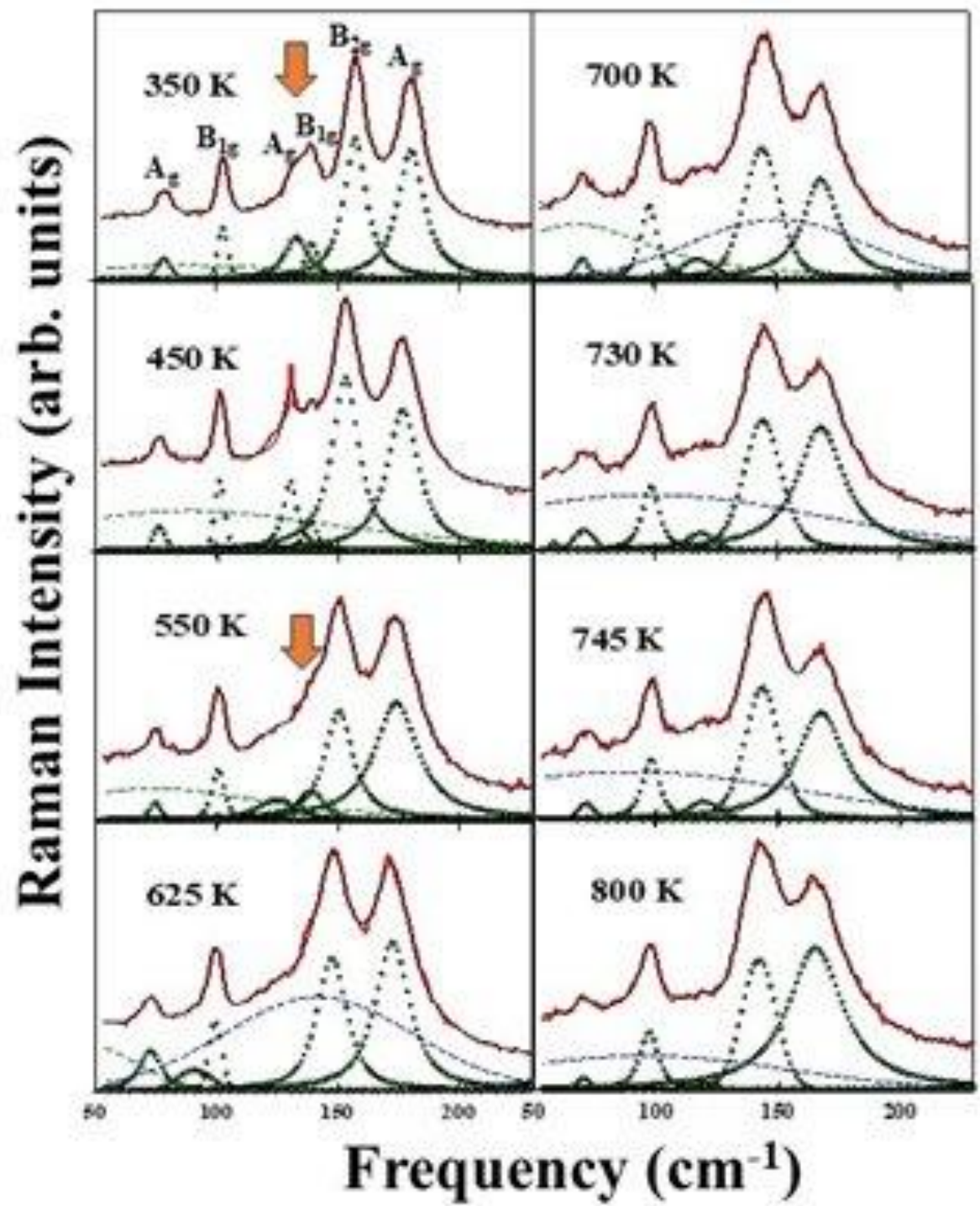

Fig. 3
MASSA et al

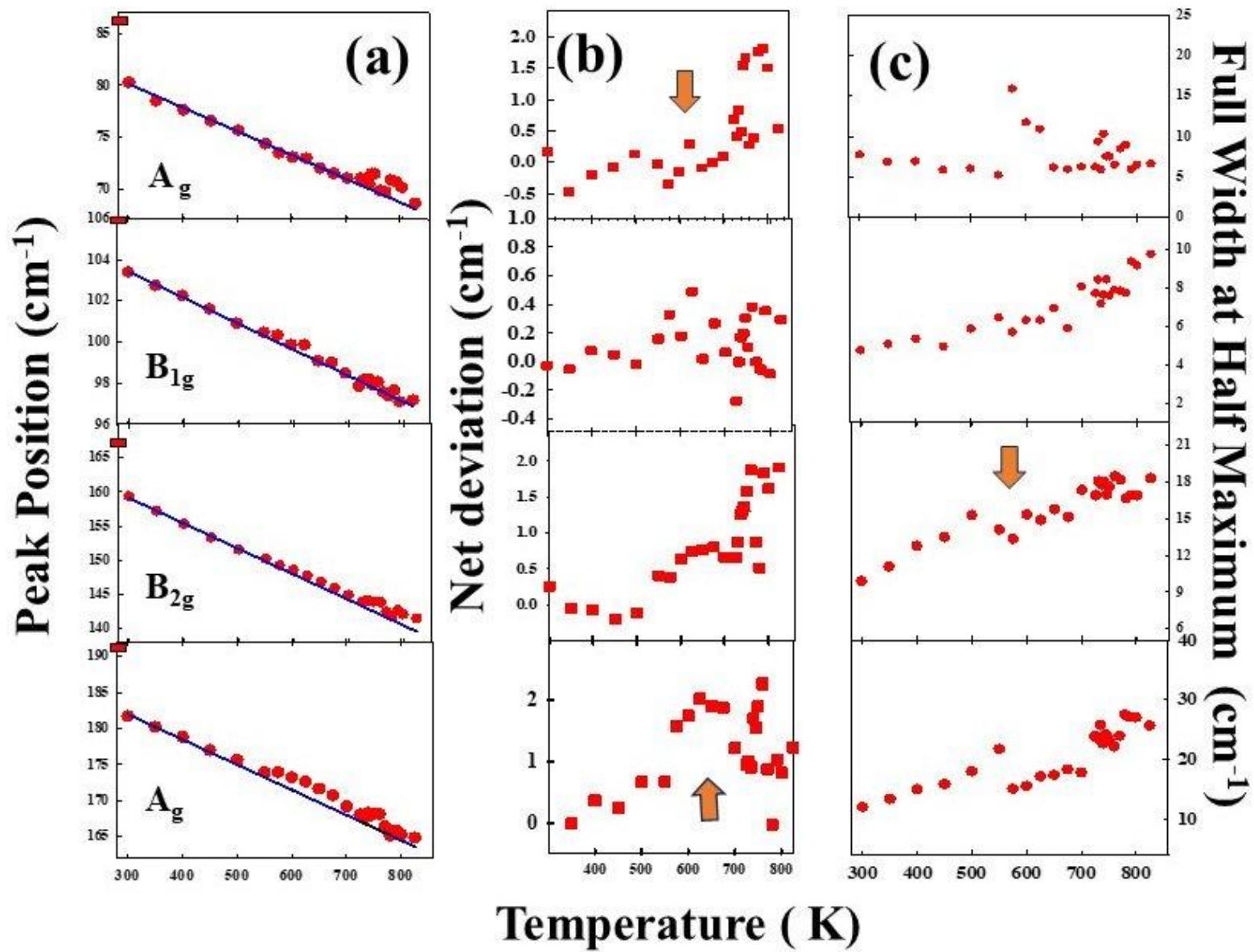

Fig. 4
MASSA et al

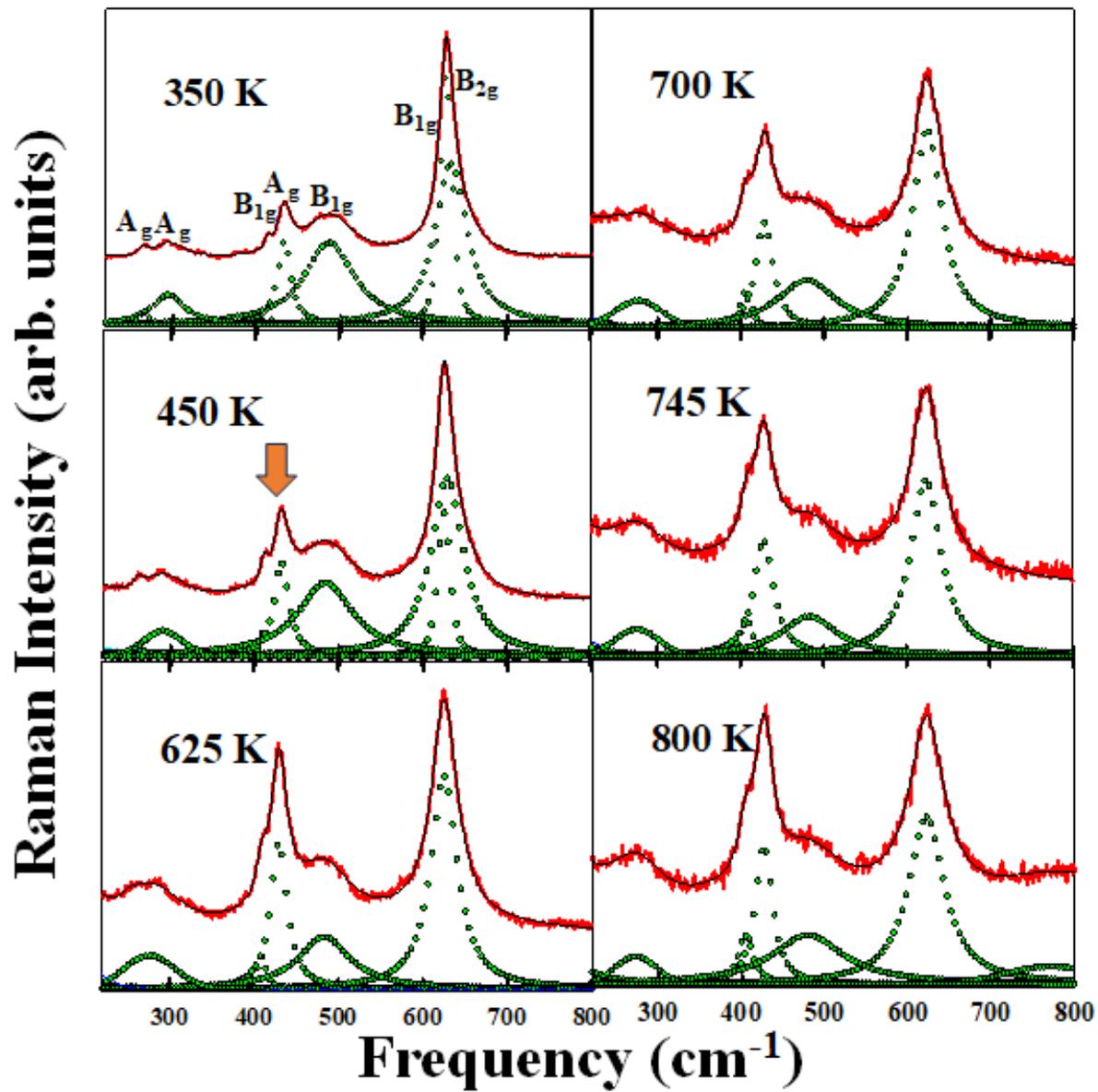

Fig. 5
MASSA et al

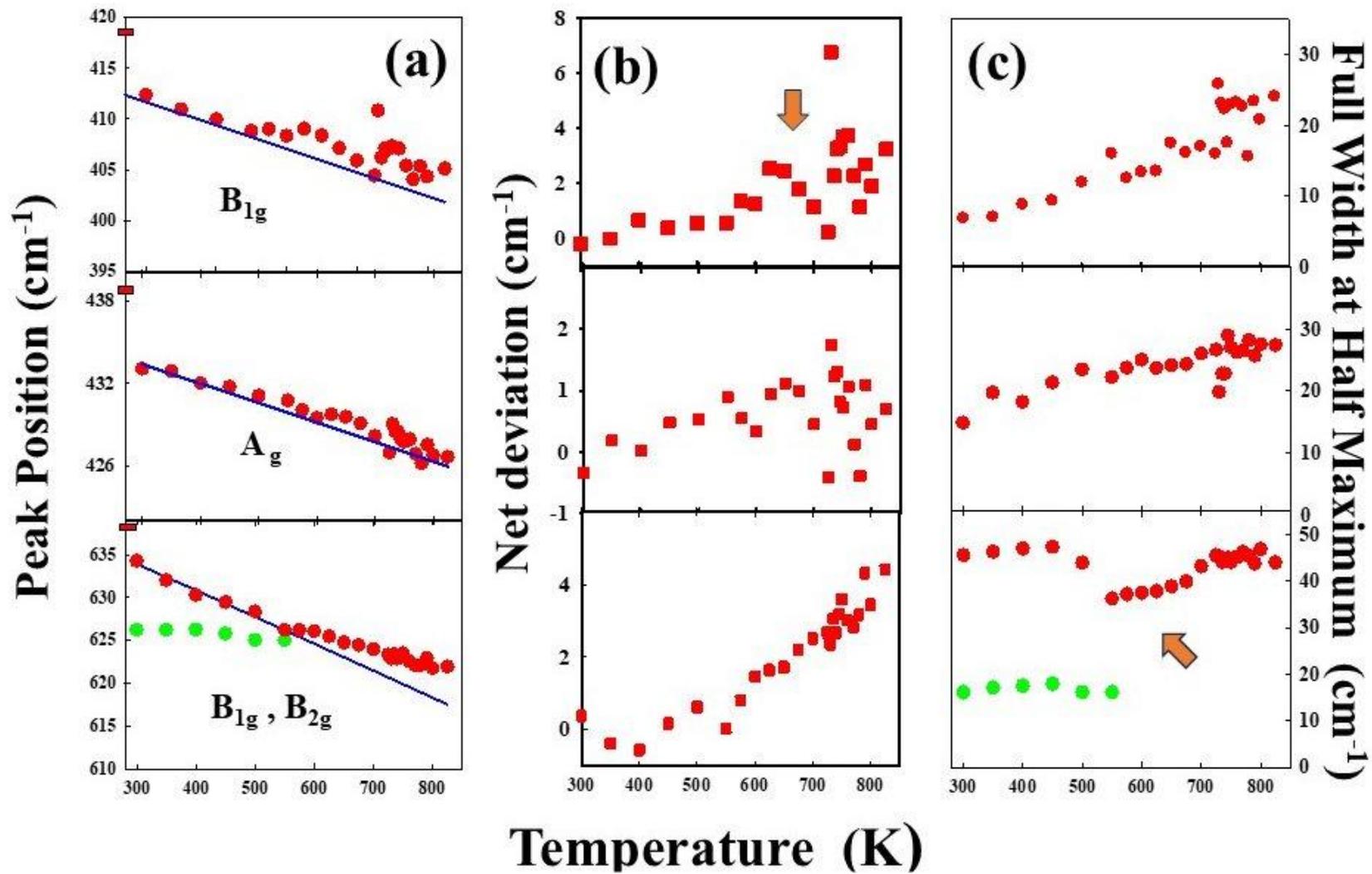

Fig. 6
MASSA et al

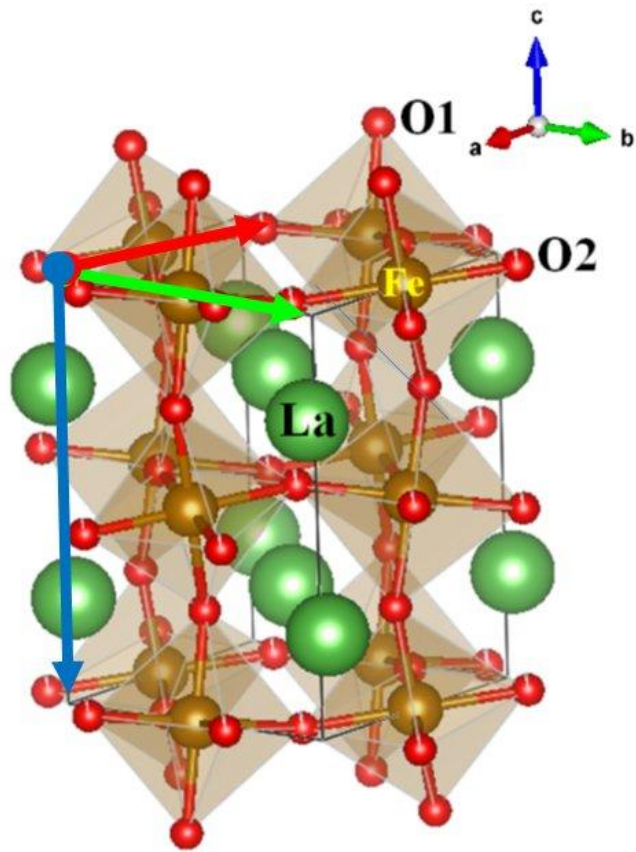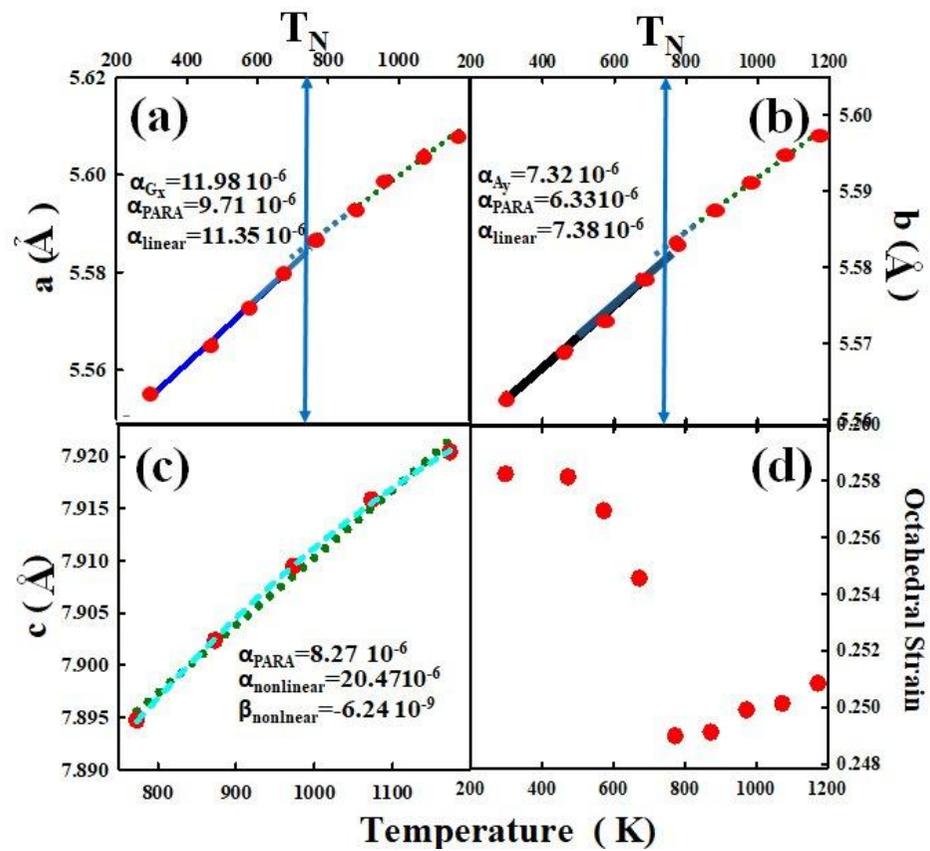

**Fig. 7**
**MASSA et al**

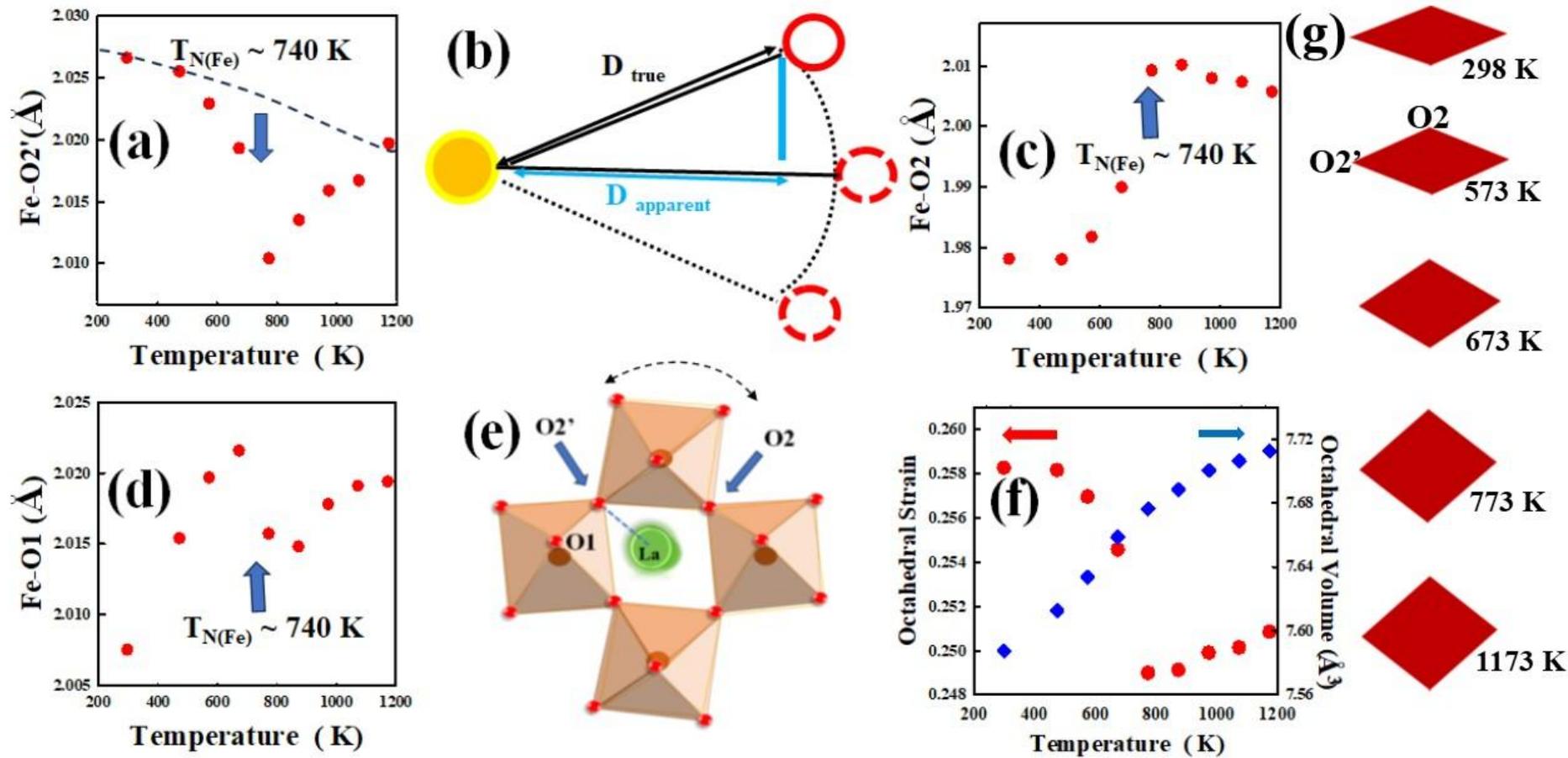

**Fig. 8
MASSA et al**

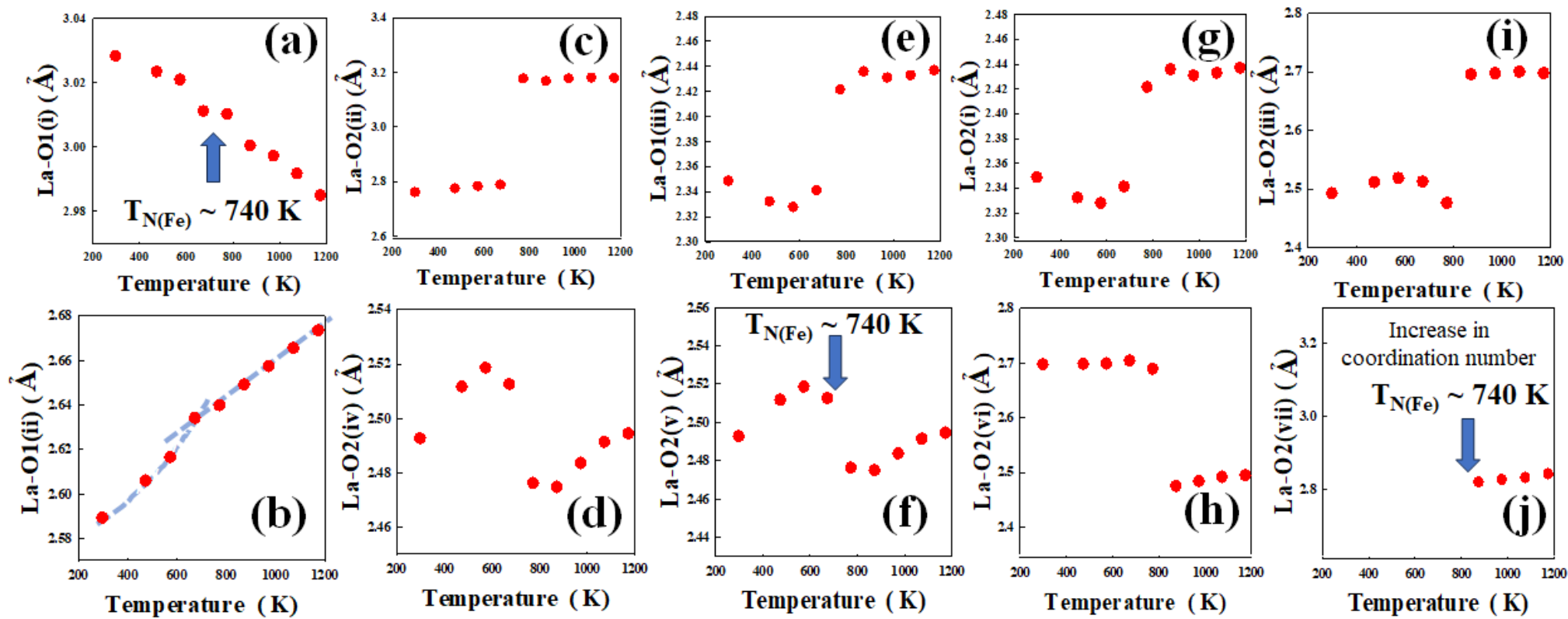

**Fig. 9
MASSA et al**

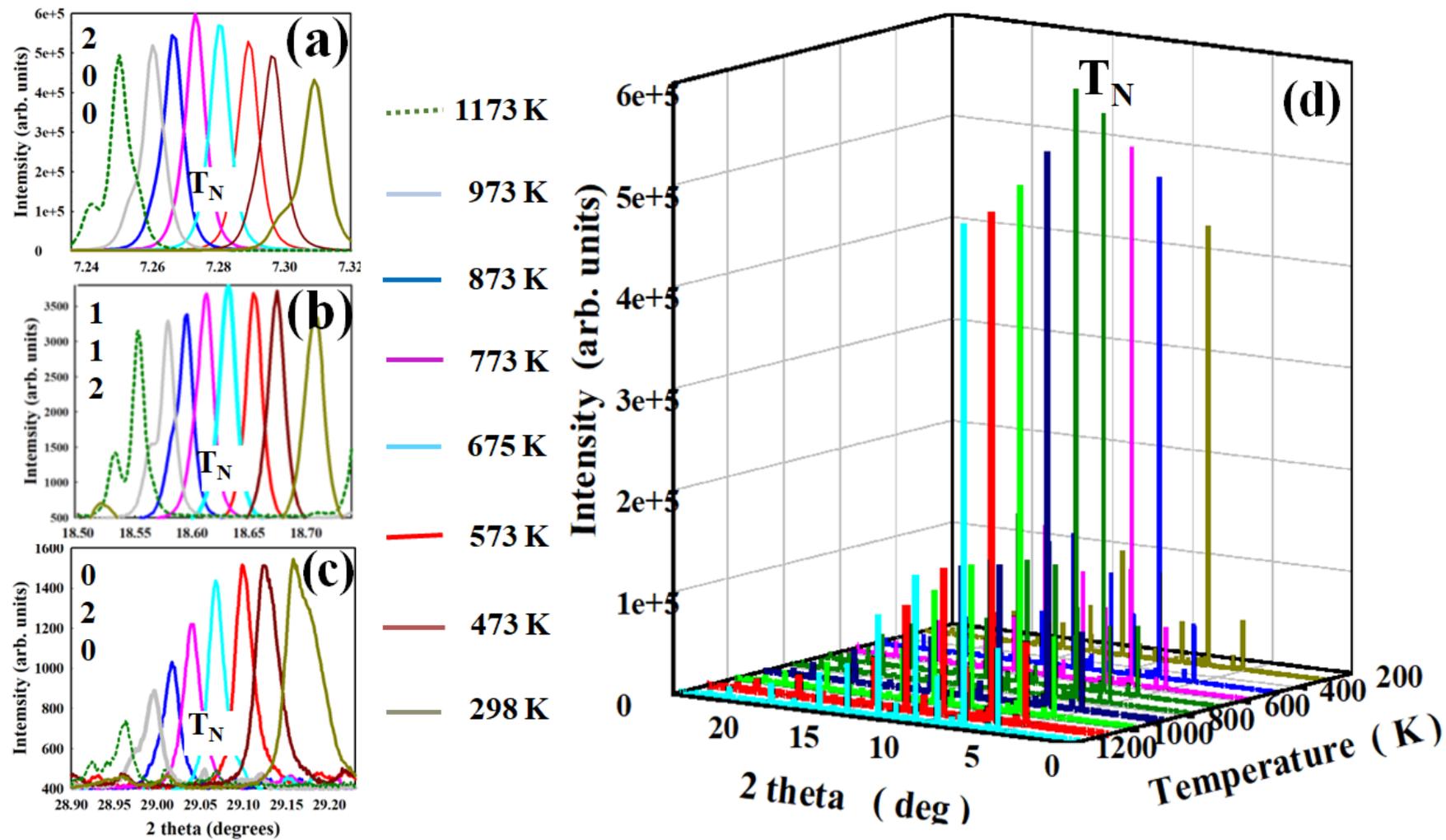

**Fig. 10**
**MASSA et al**

# SUPPLEMENTAL MATERIAL

## Spin-phonon interaction revisited: far infrared emission, Raman scattering, and high-resolution X-ray diffraction at the Néel temperature in LaFeO$_3$

Néstor E. Massa,*,1  Javier Gainza [2]  Aurélien Canizarès,[3] Leire del Campo,[3]

and

José Antonio Alonso.[4]

[1] Centro CEQUINOR, Consejo Nacional de Investigaciones Científicas y Técnicas, Universidad Nacional de La Plata, Blvd. 120 1465, B1904 La Plata, Argentina

[2] European Synchrotron Radiation Facility (ESRF), 71 Avenue des Martyrs, 38000 Grenoble, France.

[3] Centre National de la Recherche Scientifique, CEMHTI UPR3079, Université Orléans, F-45071 Orléans, France

[4] Instituto de Ciencia de Materiales de Madrid, CSIC, Cantoblanco, E-28049 Madrid, Spain

:Corresponding author:

*Néstor E. Massa,  e-mail: neemmassa@gmail.com

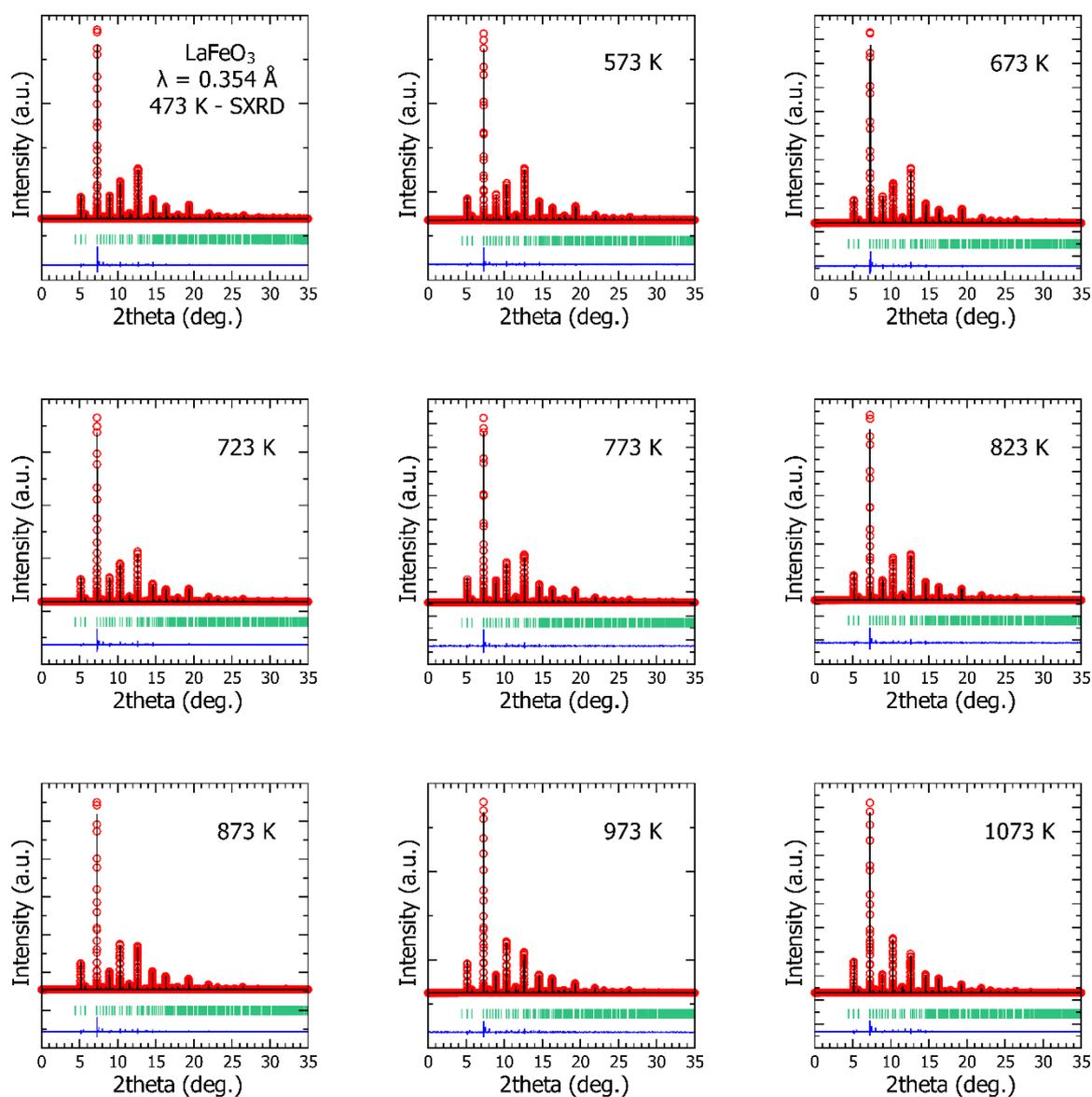

**Figure S1** Temperature-dependent on-warming X-ray diffraction patterns for LaFeO$_3$ bettween 473 K and 1073 K in the space group Pbnm (D$_{2h}^{16}$), $Z = 4$. Sharp peaks indicate very good polycrystalline quality. In the Rietveld plot, the red points correspond to the experimental profile and the solid line is the calculated profile, with the differences in the bottom blue line.

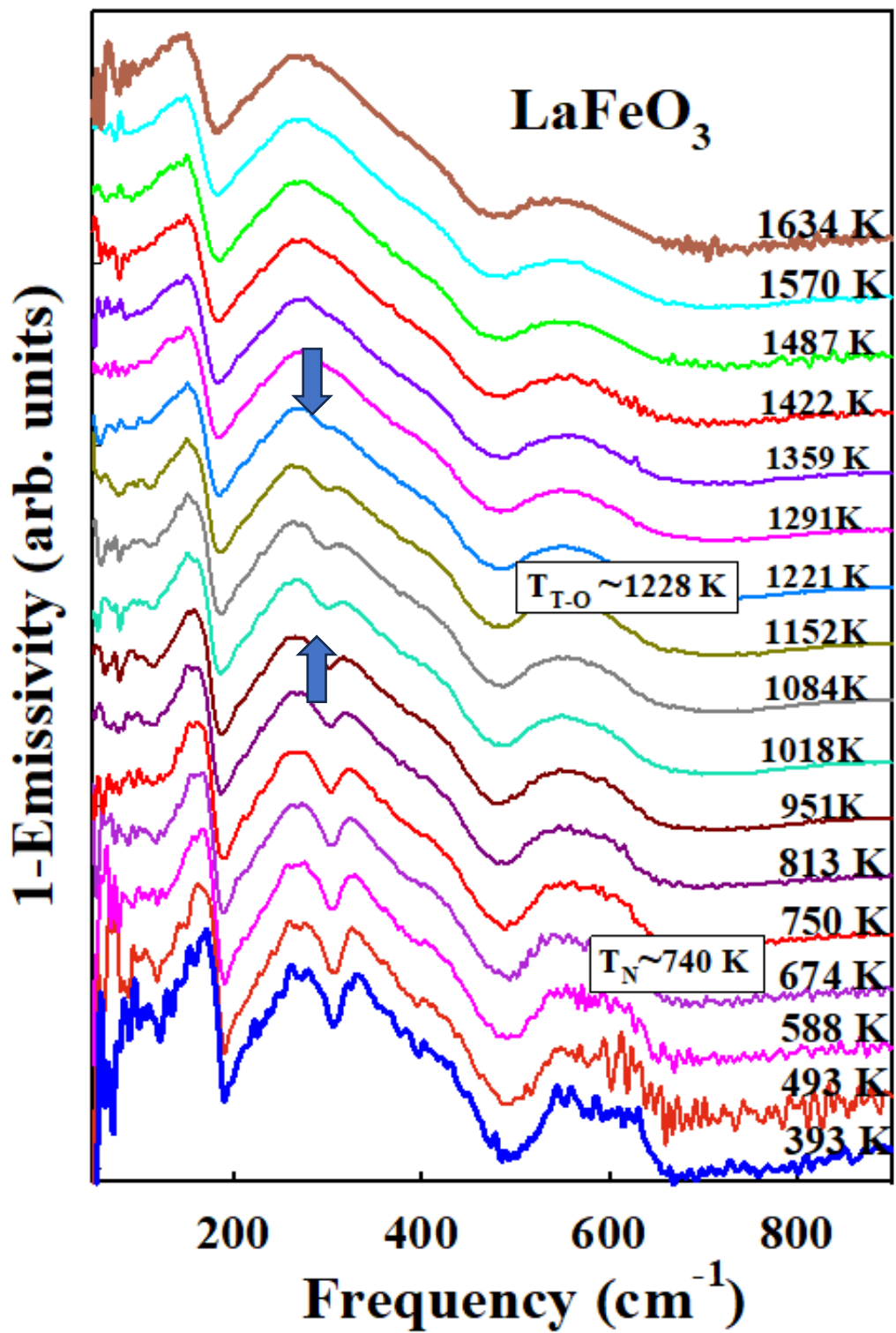

**Figure S1** LaFeO$_3$ 1-Emissivity phonon spectra from 393 K to 1634 K. Note that the orthorhombic-rhombohedral order-disorder phase transition at 1228 K, discussed elsewhere, appears (arrows) as phonon merging at about that temperature, The spectra have been vertically offset for better viewing.

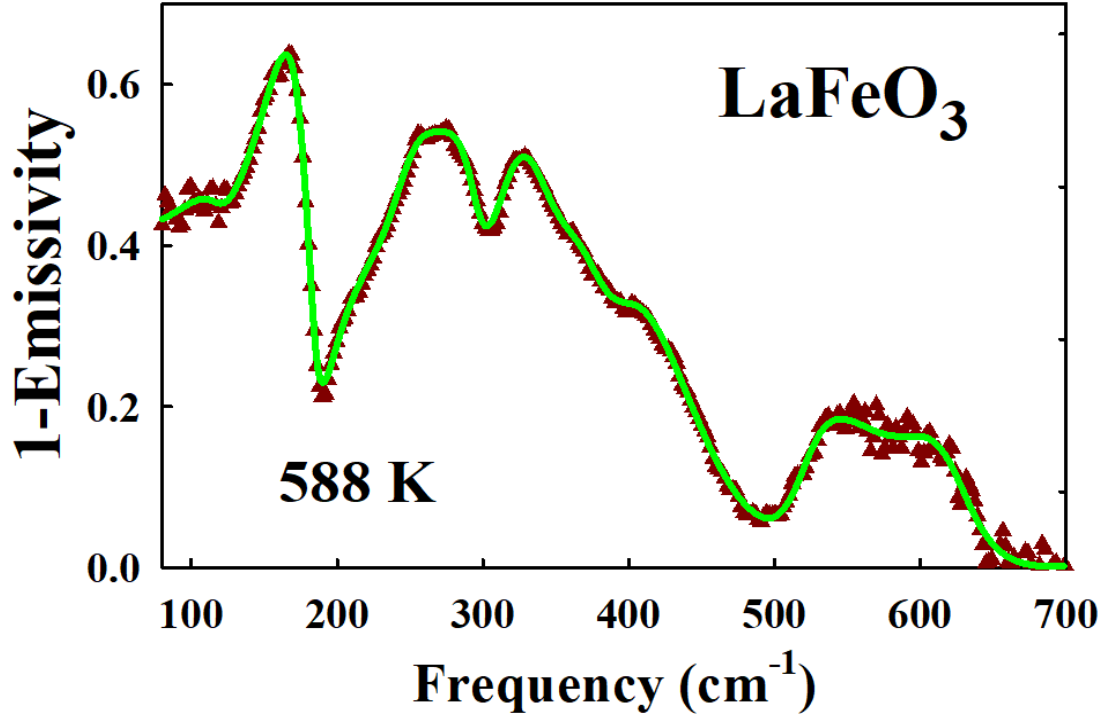

**Figure S2** LaFeO₃ near normal reflectivity at 588 K, experimental: triangles, full line: fit.

## Table S1

Dielectric simulation fitting parameters for LaFeO₃ near normal reflectivity at 588 K.

| T (K) | $\varepsilon_\infty$ | $\omega_{TO}$ (cm⁻¹) | $\Gamma_{TO}$ (cm⁻¹) | $\omega_{LO}$ (cm⁻¹) | $\Gamma_{LO}$ (cm⁻¹) |
|---|---|---|---|---|---|
| 588 | 1.45 | 72.9 | 203,9 | 12.2 | 187.6 |
| | | 118.0 | 39.4 | 3.27 | 31.9 |
| | | 157.6 | 36.3 | 17.6 | 21.8 |
| | | 175.6 | 26.3 | 9.0 | 16.0 |
| | | 215.7 | 90.7 | 21.5. | 61.7 |
| | | 248.0 | 37.9 | 18.7 | 52.9 |
| | | 275.1 | 58.1 | 22.2 | 27.9 |
| | | 306.2 | 58.2 | 3.0 | 97.9 |
| | | 316.6 | 56.1 | 31.2 | 56.2 |
| | | 350.7 | 78.7 | 31.2 | 46.8 |
| | | 384,5 | 60.0 | 10.6 | 70.0 |
| | | 397.9 | 58.6 | 39.3 | 11.1 |
| | | 444.8 | 208.7 | 41.5 | 70.4 |
| | | 484.5 | 56.7 | 3.9 | 75.7 |
| | | 521.5 | 41.31 | 6.2 | 109.3 |
| | | 546.1 | 116.4 | 34.1 | 113.1 |
| | | 602.4 | 70.7 | 30.5 | 51.2 |
| | | 1421.6 | 796.3 | 865.2 | 2453.3 |

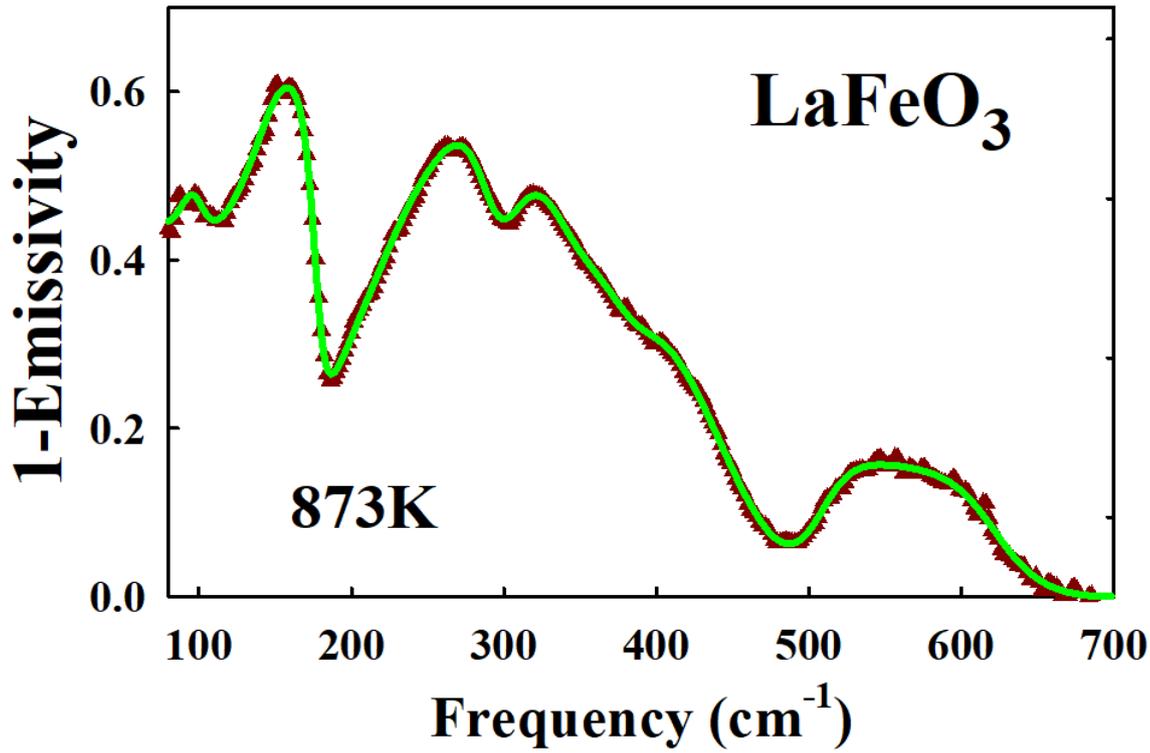

**Figure S3** LaFeO3 near normal reflectivity at 873 K, experimental: triangles, full line: fit.

**Table S2**
Dielectric simulation fitting parameters for LaFeO3 near normal reflectivity at 873 K

| T (K) | $\varepsilon_\infty$ | $\omega_{TO}$ (cm$^{-1}$) | $\Gamma_{TO}$ (cm$^{-1}$) | $\omega_{LO}$ (cm$^{-1}$) | $\Gamma_{LO}$ (cm$^{-1}$) |
|---|---|---|---|---|---|
| 873 | 1.47 | 68.3 | 127.1 | 14.5 | 172.5- |
| | | 96.0 | 19.1 | 3.2 | 13.2 |
| | | 147.6 | 51.5 | 22.0 | 61.6 |
| | | 173,9 | 43.5 | 4.70 | 18.3 |
| | | 205.2 | 220..9 | 20.1 | .139.4 |
| | | 231.0 | 70.0 | 19.1 | 108.7 |
| | | 269.6 | 71.3 | 22.5 | 44.5 |
| | | 298.5 | 252.1 | 3.6 | 116.2 |
| | | 304.6 | 46.5 | 31.2 | 96.7 |
| | | 347.7 | 151.3. | 31.6 | 54.7 |
| | | 378.0 | 66.7 | 10.7 | 93.6 |
| | | 392.2 | 71.2 | 38.7 | 104.9 |
| | | 445.1 | 224.5 | 39.4 | 113.3 |
| | | 478.5 | 155.0 | 3.9 | 105.3 |
| | | 509.8 | 51.8 | 34.5 | 162.0 |
| | | 594,5 | 209.4 | 30.5 | 56.8 |
| | | 1284.98 | 1425.2 | 1138.6 | 1402.8 |

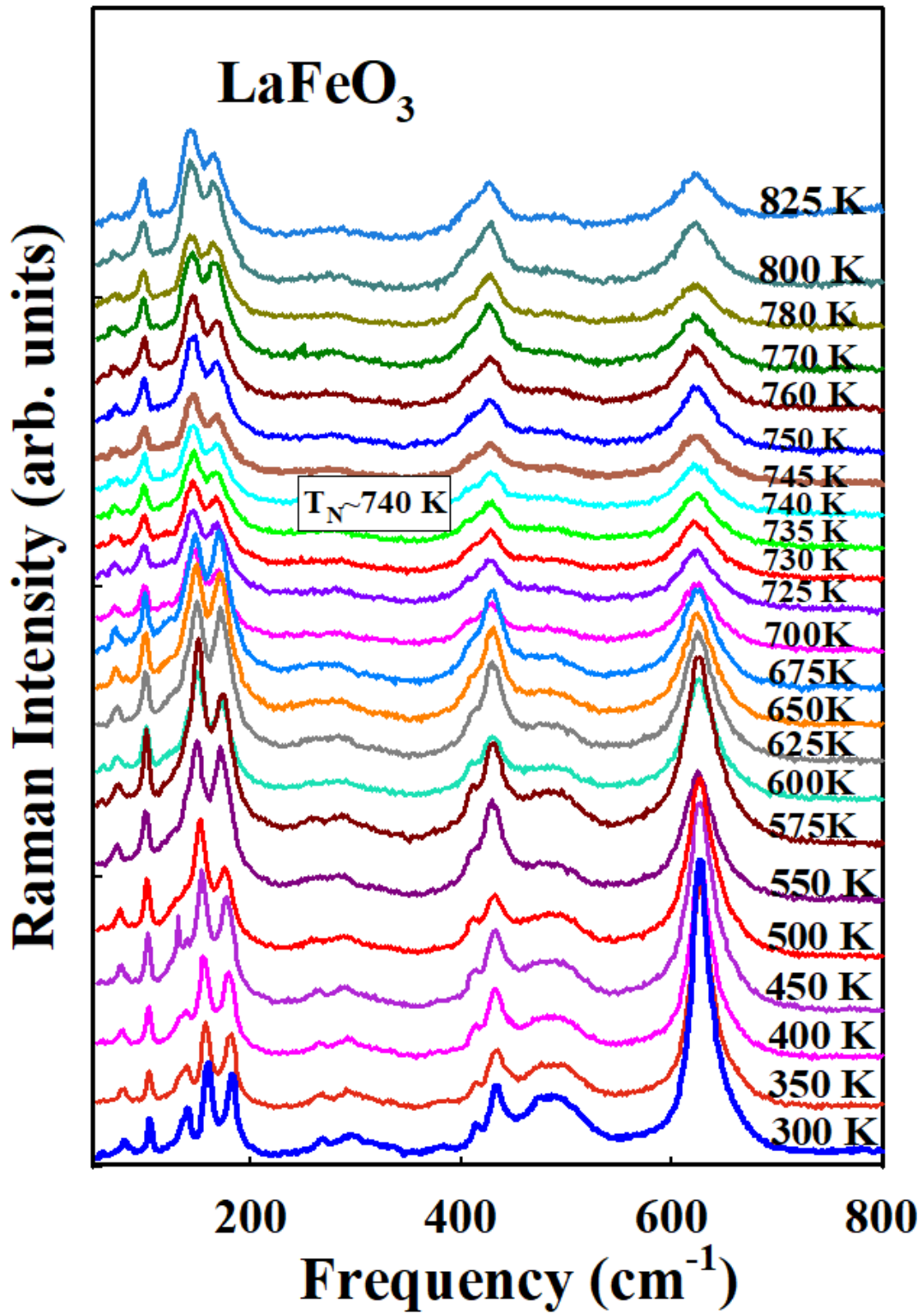

**Figure S4** Temperature-dependent LaFeO₃ Raman spectra in the 300 K to 800 K range using laser line λexc = 633 nm. The spectra have been vertically offset for better viewing

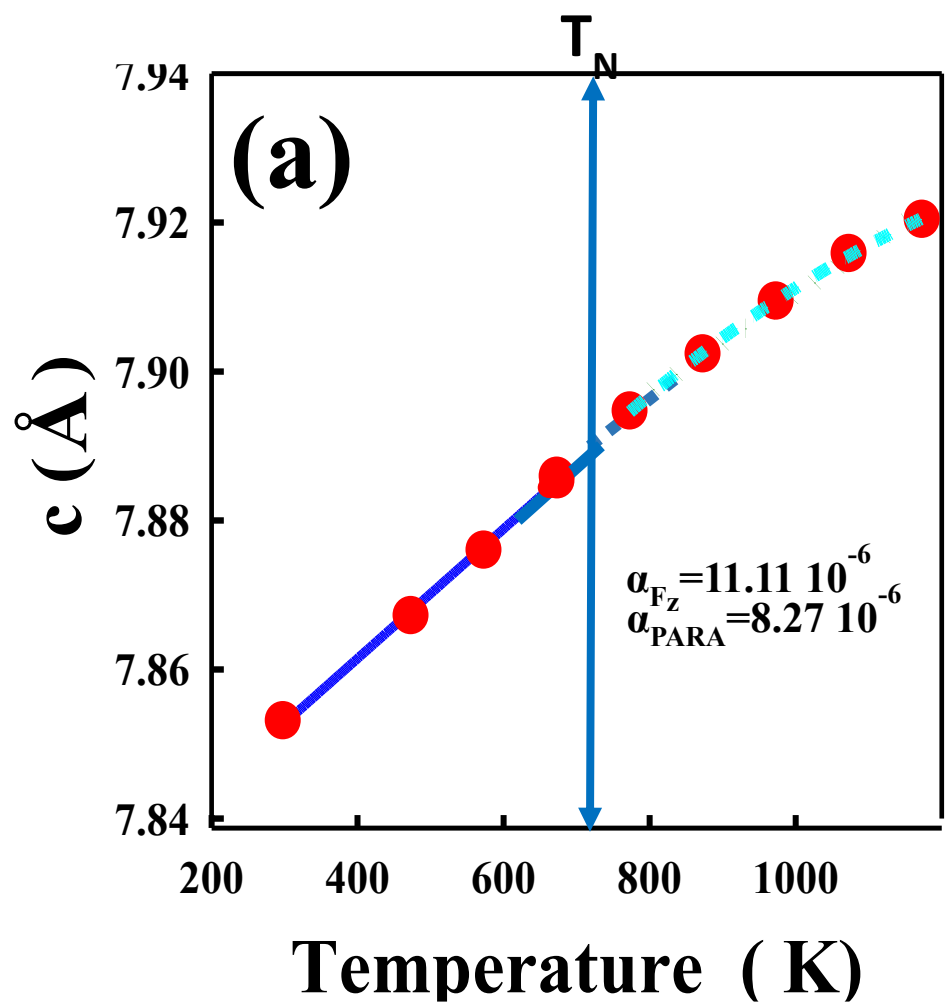 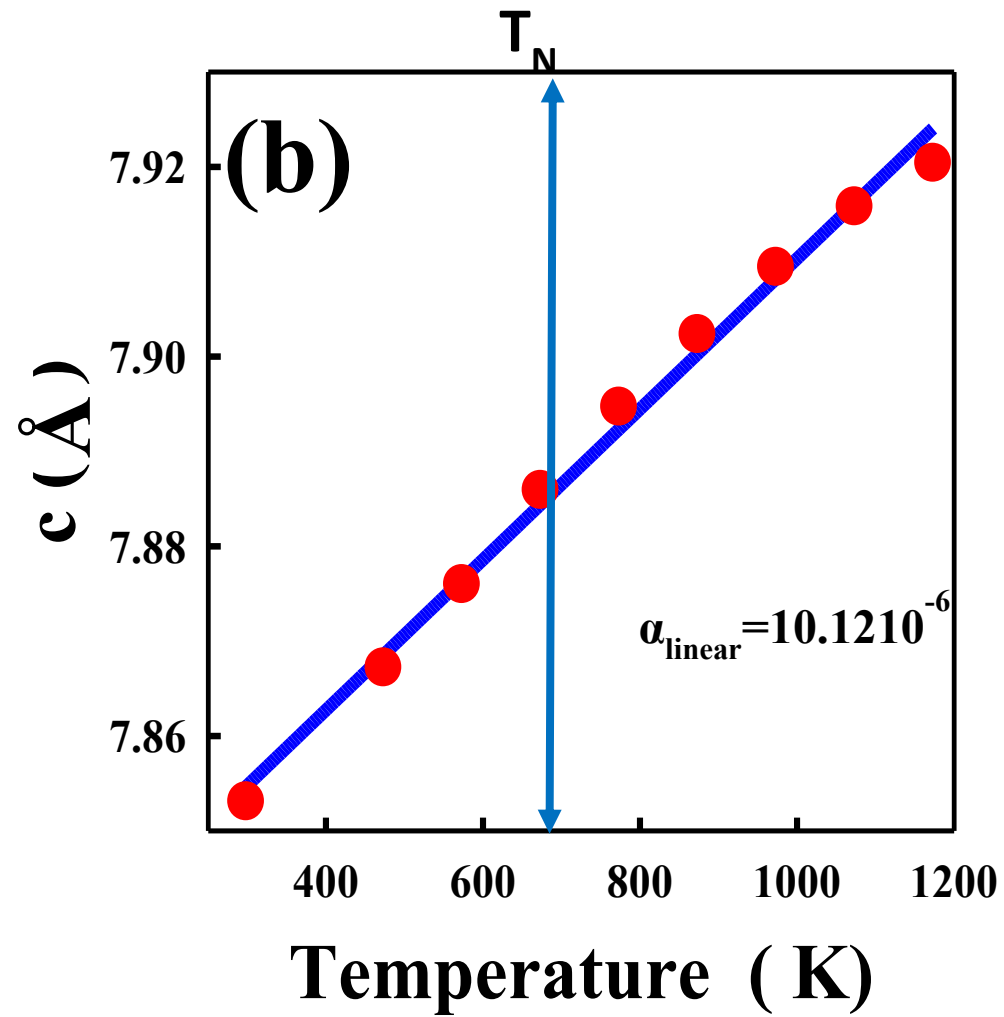

Fig S 5(a) Temperature dependent lattice constant $\underline{c}$ determined from measured atomic positions in the space group Pbnm $(D_{2h}^{16})$; linear (dotted line) and quadratic non-linear (dashed line) fit. Inset : $\alpha_{Fz}$ (K$^{-1}$) linear thermal expansion coefficient in the noncolineal ferromagnetic phase, $\alpha_{PARA}$ (K$^{-1}$) linear thermal expansion coefficient in the paramagnetic phase (b) Linear fit to lattice constant $\underline{c}$ in the complete temperature range , inset: $\alpha_{linear}$ (K$^{-1}$) linear thermal expansion coefficient in the full temperature range,

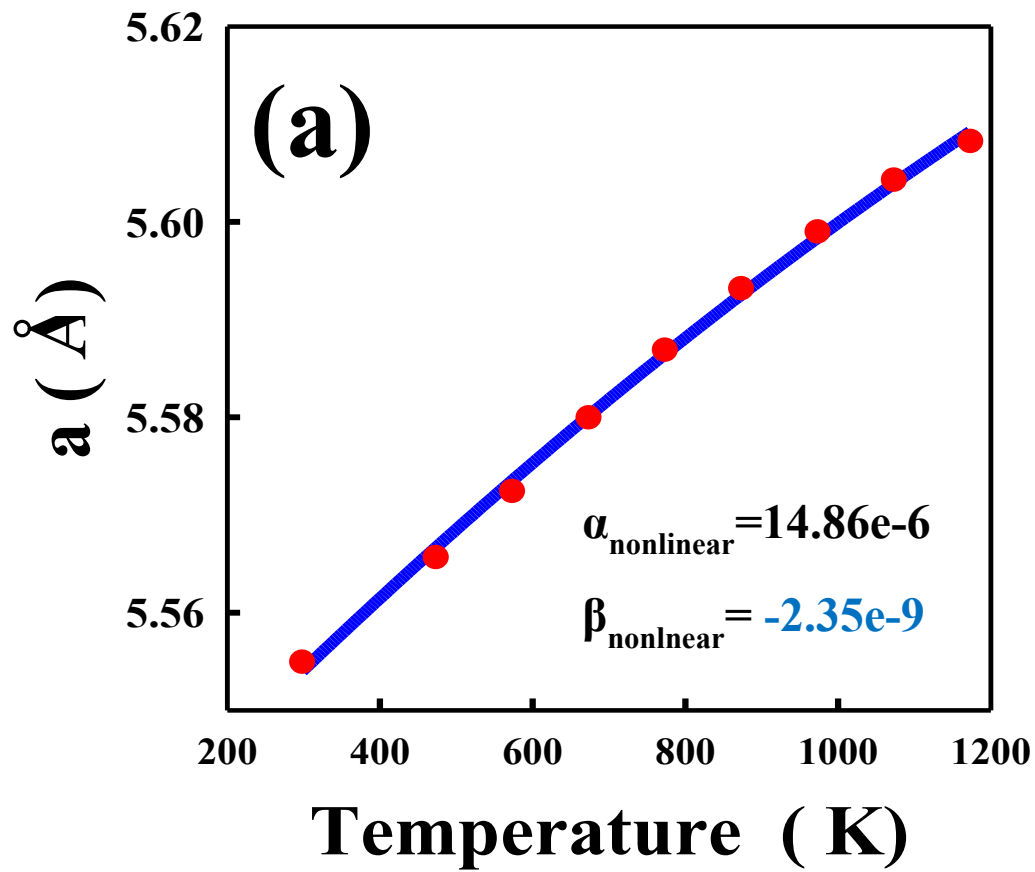 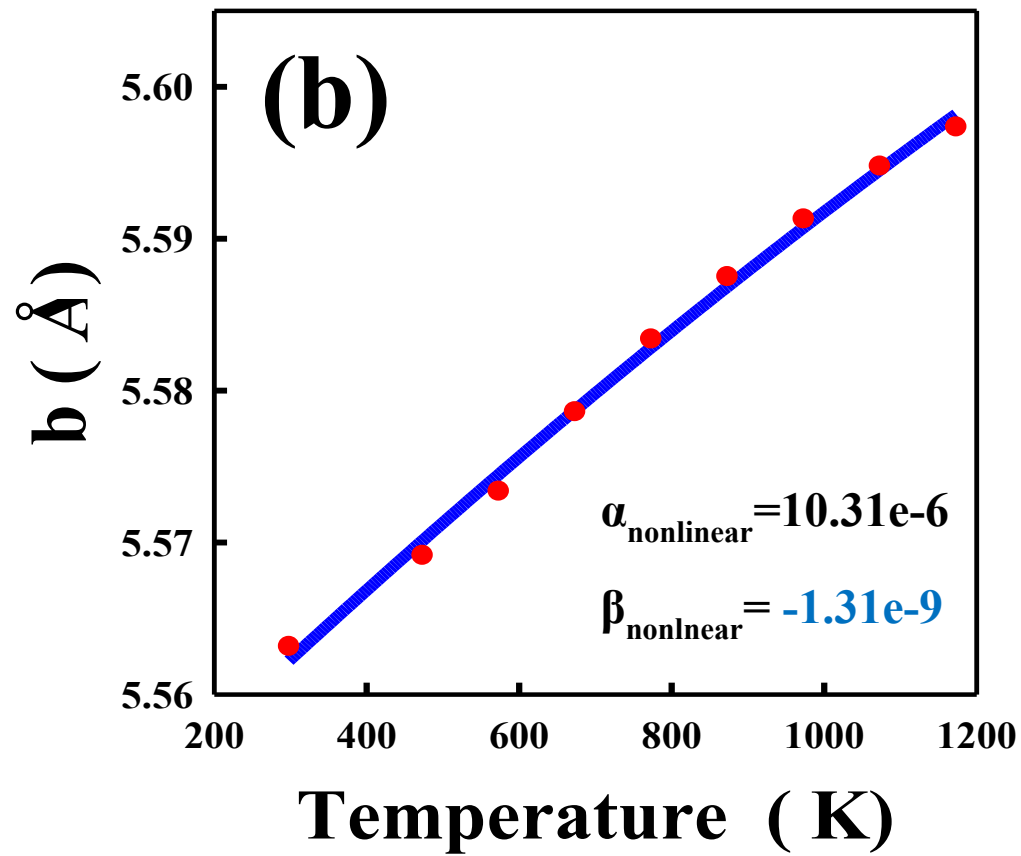

Fig. S 6 (a) Nonlinear fit to lattice constant *a* determined from measured temperature-dependent atomic positions in the space group Pbnm ($D_{2h}^{16}$) .inset: $α_{nonlinear}$: ($K^{-1}$) Linear thermal expansion term calculated from the non-linear fit of the lattice constant *c* , $β_{nonlnear}$ ($K^{-2}$) : Non-linear thermal expansion term calculated from the non-linear fit to the lattice constant *a* ; (b) Nonlinear fit to lattice constant *b* determined from measured temperature-dependent atomic positions in the space group Pbnm ($D_{2h}^{16}$); inset: $α_{nonlinea}$ ($K^{-1}$) $_r$:Linear thermal expansion term calculated from the non-linear fit to the lattice constant *c* , $β_{nonlnea}$ ($K^{-2}$) $_r$: Non-linear thermal expansion term calculated from the non-linear fit to the lattice constant *b*.

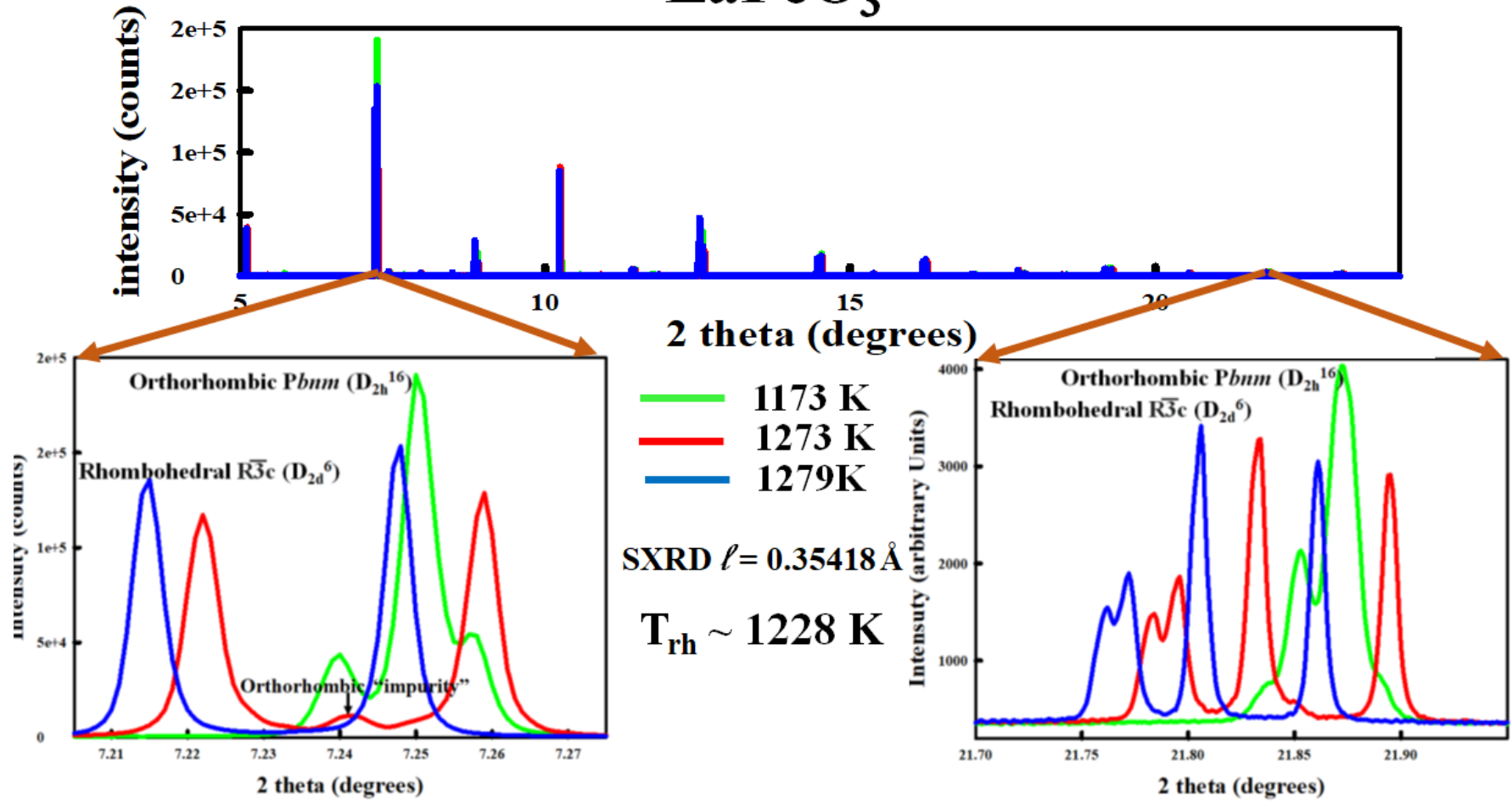

**Figure S7** Preliminary data of the temperature-dependent on-warming gradual structural order-disorder phase transition in LaFeO$_3$ from orthorhombic to rhombohedral around nominal T$_{Rh}$~ 1228 K

# Apex Oxygen O1 ↔ La and Apex Oxygen O1 ↔ Fe distance

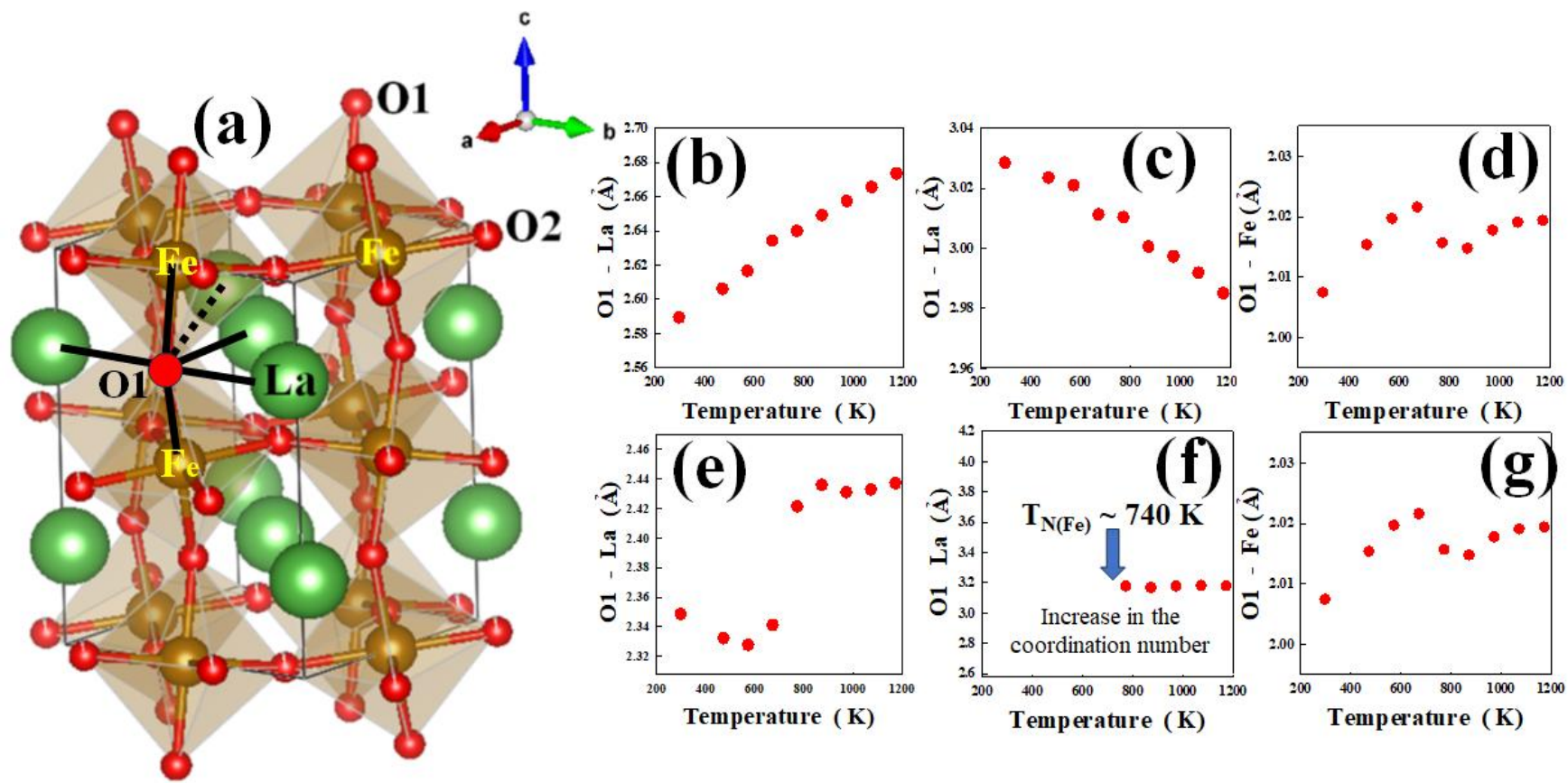

**Figure S8** Temperature-dependent "O1" distances; O1-La and O1-Fe nearest neighbor bond lengths.; (f) Note that the O1-La is only considered upon the relative increase in symmetry in the paramagnetic phase

# Oxygen basal O2 ↔ La , Oxygen basal O2 ↔ Fe distance

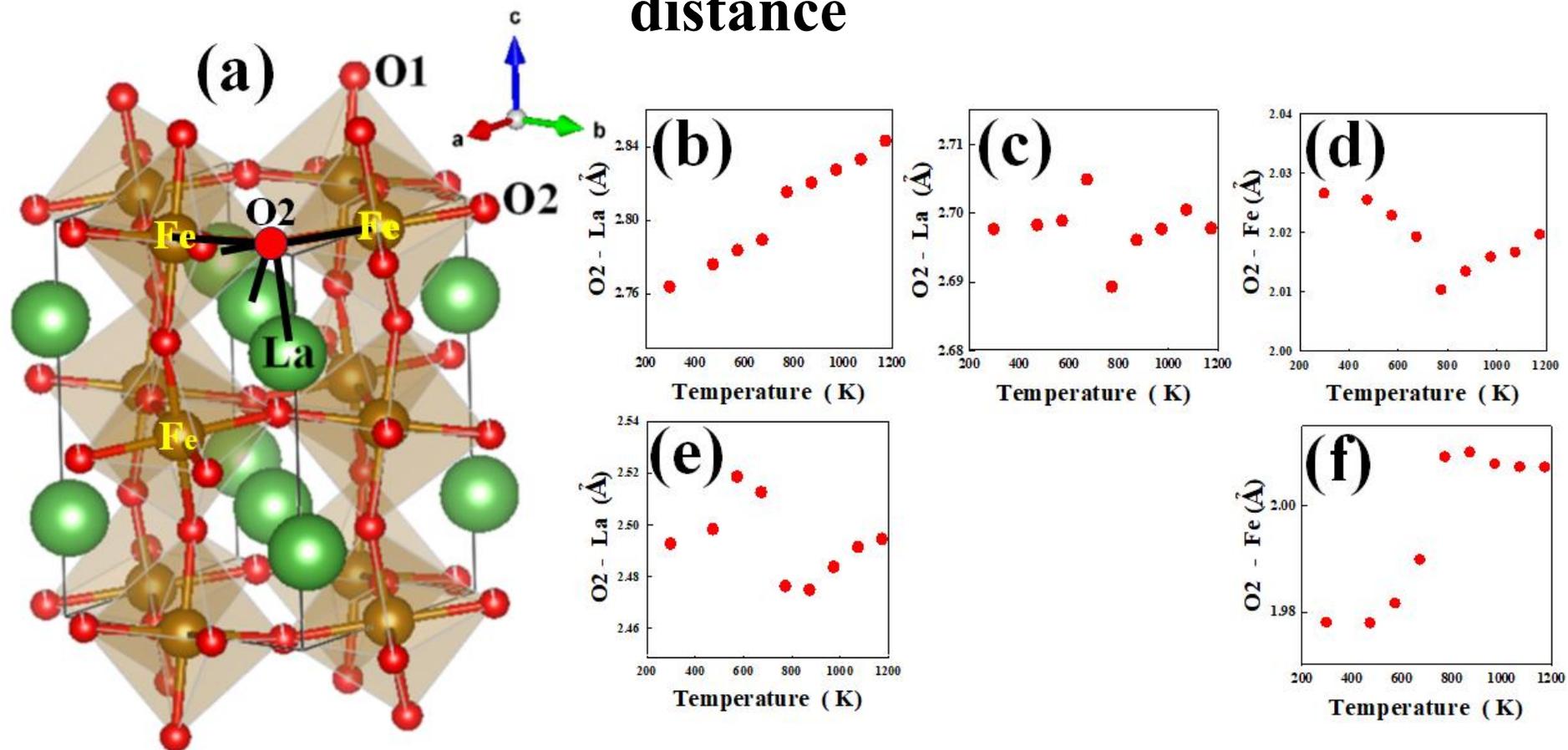

**Figure S9** Temperature-dependent "O2" distances, O2-La and O2-Fe nearest neighbor bond lengths.